\documentclass[preprint,12pt]{elsarticle}

\usepackage{amssymb}
\usepackage{amsmath}

\usepackage{svg}
\usepackage{subcaption}
\usepackage{multirow}
\usepackage{graphicx}

\journal{Signal Processing: Image Communication}

\begin{document}

\begin{frontmatter}



\title{Lightweight Super Resolution-enabled Coding Model for the JPEG Pleno Learning-based Point Cloud Coding Standard}


\author[label1]{André F. R. Guarda} 
\author[label1,label2]{Nuno M. M. Rodrigues\fnref{fn1}}
\author[label1,label3]{Fernando Pereira}

\fntext[fn1]{Nuno M. M. Rodrigues is an EURASIP member.}

\affiliation[label1]{organization={Instituto de Telecomunicações},
            city={Lisbon},
            postcode={1049-001}, 
            country={Portugal}}
\affiliation[label2]{organization={ESTG, Politécnico de Leiria},
            city={Leiria},
            postcode={2411-901}, 
            country={Portugal}}
\affiliation[label3]{organization={Instituto Superior Técnico, Universidade de Lisboa},
            city={Lisbon},
            postcode={1049-001}, 
            country={Portugal}}

\begin{abstract}
While point cloud-based applications are gaining traction due to their ability to provide rich and immersive experiences, they critically need efficient coding solutions due to the large volume of data involved, often many millions of points per object. The JPEG Pleno Learning-based Point Cloud Coding standard, as the first learning-based coding standard for static point clouds, has set a foundational framework with very competitive compression performance regarding the relevant conventional and learning-based alternative point cloud coding solutions. This paper proposes a novel lightweight point cloud geometry coding model that significantly reduces the complexity of the standard, which is essential for the broad adoption of this coding standard, particularly in resource-constrained environments, while simultaneously achieving small average compression efficiency benefits. The novel coding model is based on the pioneering adoption of a compressed domain approach for the super-resolution model, in addition to a major reduction of the number of latent channels. A reduction of approximately 70\% in the total number of model parameters is achieved while simultaneously offering slight average compression performance gains for the JPEG Pleno Point Cloud coding dataset.
\end{abstract}




\end{frontmatter}




\section{Introduction}
\label{sec1}

Immersion is a fundamental characteristic of real-world visual experiences. It allows the users to move around, rotating and translating in space, and instantaneously perceive the light arriving from new angles and positions. This capability is essential for multiple emerging application scenarios, notably virtual, augmented, and mixed reality as well as cultural heritage, medical, education, architecture, and engineering. Beyond human experiences, visual immersion is becoming increasingly crucial for machines, particularly autonomous vehicles and collaborative robots, making it a largely required capability also for non-human consumption.

Meaningful immersive experiences demand powerful 3D representation models, able to create rich 3D visual representations by describing the full light rays in a scene, namely the intensity of light at each position ($x,y,z$), for any viewing angle ($\theta,\phi$), at any time ($t$) and any wavelength ($\lambda$). This led to the definition of the so-called 7D plenoptic function P($x,y,z,\theta,\phi,t,\lambda$) \cite{Adelson91} which became a reference for how rich and complete 3D visual representations can be. The most common 3D representation models targeting to represent the full plenoptic function data are light fields, point clouds and meshes.

To offer rich experiences, the acquired visual data must be complete and detailed enough, thus asking for spatial, temporal and angular resolutions and precisions which lead to massive amounts of raw data. In this context, the practical deployment of plenoptic data-based storage and streaming applications critically needs efficient coding solutions to significantly reduce the rates for the target quality of experience. This has led the Joint Photographic Experts Group (JPEG) to launch, around 2015, the JPEG Pleno project, targeting the specification of coding standards for plenoptic data modalities \cite{Ebrahimi06,Astola20}.

From the three representation approaches mentioned above, this paper focuses on point clouds which are gaining relevance due to the growing availability of acquisition sensors, and their versatility, easy interactivity and navigation, as well as lightweight nature, which supports real-time processing. A point cloud (PC) consists of an unordered set of points in 3D space representing the surface of an object or scene, which may be static or dynamic, i.e., varying in time. While the ($x, y, z$) 3D spatial coordinates of the points define the so-called PC geometry, each point may have associated one or more attributes such as color, reflectance, or normal vectors, among others, to enhance the PC representation and its associated immersive capabilities. Like other plenoptic modalities, PCs critically require efficient compression, as high-quality experiences often demand PCs with many millions of points.

As demonstrated by other modalities in the past, e.g., images and video, PC coding (PCC) standardization is crucial to offer the interoperability required by many applications. To address this goal, the Moving Picture Experts Group (MPEG) specified two key PCC standards in the early 2020s: Geometry-based Point Cloud Compression (G-PCC) for static PCs and Video-based Point Cloud Compression (V-PCC) for dynamic PCs \cite{Graziosi20,Cao21}. These standards evolved over time to improve compression efficiency and are commonly referred to as conventional since largely relying on handcrafted coding techniques, like those used in media coding for the past three decades.

However, around 2018, artificial intelligence (AI) emerged in the visual coding arena, after major impacts on computer vision tasks such as recognition and classification. This shift coincided with the exponential increase in machine-driven visual data consumption. Learning-based visual data coding initially focused on modalities like 2D images \cite{Toderici16,Balle18,Minnen18}, a very structured modality since the pixels fully fill a regular grid, but soon expanded to 3D data, notably PCs, an unstructured modality since the points sparsely and irregularly fill a 3D grid. 

Driven by this evolution, in 2022, JPEG launched two learning-based coding standards: JPEG AI for 2D image coding \cite{JpegAIcfp,JpegAIresults,JpegAIstandard,Ascenso23} and JPEG Pleno Learning-based Point Cloud Coding (JPEG PCC) for static PCs. The JPEG PCC Call for Proposals \cite{JpegPCCcfp}, issued in January 2022, set an ambitious goal: to develop learning-based coding for static PC geometry and their attributes while ensuring high efficiency for both human visualization and machine processing. The aim was to provide competitive compression performance compared to existing PCC standards while establishing a royalty-free baseline. This was a rather revolutionary goal since, for the first time, a single compressed representation should efficiently and effectively address the needs of man and machine consumption, unifying the two worlds which had traditionally been treated separately.

The first version of the JPEG PCC standard, finalized in 2024, incorporates both geometry and texture (color) coding \cite{JpegPCCstandard,Guarda25}. While geometry coding employs a newly designed deep-learning (DL) model, texture coding leverages the new JPEG AI standard \cite{JpegAIstandard,Ascenso23}, after projecting the 3D textures into 2D images for JPEG AI coding. This results in a JPEG-driven DL-based coding framework for both images and PCs, which JPEG plans to extend to other modalities, e.g., light fields and event data. The JPEG PCC geometry coding architecture includes a DL-based geometry representation model, consisting of separate coding and super-resolution (SR) models; it also includes other modules such as block partitioning, block down-sampling and voxel binarization \cite{JpegPCCstandard,Guarda25}. In fact, to achieve high compression efficiency for sparse PCs and reach lower rates for all PCs, it is convenient to subsample the original PC with a so-called \textit{sampling factor} (\textit{SF}) which commonly takes the values 1 (i.e., no down-sampling is performed), 2 or 4 (i.e., the number of voxels is reduced by a factor of 2 or 4 in the three spatial directions). The SR model is independently applied to the PC decoded by the coding model, meaning it acts on the PC in the decompressed domain.

In terms of compression performance, JPEG PCC performs considerably well for geometry coding, notably outperforming both the conventional MPEG PCC standards for most solid and dense PCs \cite{Guarda25}. Compared to other DL-based PC geometry coding solutions in the literature \cite{Wang21sparse,Liu22,Pang22}, JPEG PCC is also very consistent, achieving competitive compression performance for all types of PC content, although not necessarily the best for all PCs individually, especially when compared with coding solutions using multiple models, each tuned to a different type of PC, e.g., in terms of sparsity \cite{Guarda25}.

However, the DL-based models used in the JPEG PCC standard are quite complex in terms of the number of model parameters, and thus memory footprint, which is further aggravated considering the use of two types of DL-based models, for coding and SR, with multiple instances: for coding, there are five coding models, each with around 5.1 million parameters, targeting different rate-distortion (RD) trade-offs; for SR, there are two SR models, each with around 7.3 million parameters, for SF=2 and SF=4. This means that the SR models are heavier than the coding models, with the extra disadvantage of penalizing the decoder side where the complexity constraints are typically more critical. In addition, the coding and SR models have very different architectures and are independently trained, with the SR models not considering the effects of compression, which may hinder their performance.

The objective of this paper is to propose a novel lightweight SR-Enabled PC geometry Coding model (SR-EPCC), aiming to tackle the aforementioned shortcomings and upgrade the JPEG PCC standard to reduce its associated complexity. The proposed SR-EPCC uses a single model to address both the coding and SR functions, by adopting a low complexity compressed domain SR approach. Differently from the current JPEG PCC decompressed domain SR approach, the compressed domain approach allows the SR model to act on the features extracted from the original PC data, thus avoiding the need to perform feature extraction in the decompressed domain while also reducing the decoder side complexity. The novel SR-enabled coding model is based on a multi-branch approach, notably considering three branches for the relevant SF values. The training process, which has a critical impact on model’s performance, uses several stages to optimize the coding and compressed domain SR components of the full SR-enabled coding model. Moreover, following a study on the practical impact of the JPEG PCC latents, the baseline coding model itself is simplified, also contributing to reduce the complexity of the finally proposed coding model. In total, the proposed SR-EPCC model offers a major reduction on the overall model complexity, notably 70\% fewer model parameters, and even with small average gains in compression performance.

The integration of the novel SR-enabled coding model in the JPEG PCC standard architecture leads to the so-called \textit{JPEG SR-EPCC codec}. If adopted by JPEG, the SR-EPCC model would significantly increase the probability of this standard being used in real-world implementation platforms, due to its lower complexity requirements.

The remainder of this paper is structured as follows. Section \ref{sec2} briefly reviews the literature on conventional and learning-based PCC. Section \ref{sec3} provides a detailed overview of the JPEG PCC framework and its key components. Section \ref{sec4} introduces the proposed SR-EPCC model, while Section \ref{sec5} presents experimental results for the JPEG SR-EPCC codec in comparison to relevant benchmarks under meaningful conditions and datasets. Section \ref{sec6} concludes the paper and highlights some future work.

\section{Related Background}
\label{sec2}

This section presents a brief summary of background PCC solutions, both conventional and learning based. The current conventional PCC solutions are largely dominated by the two MPEG PCC standards which address different types of PCs \cite{Graziosi20,Cao21}. G-PCC, which is designed primarily for static and sparse PCs, directly encodes the 3D geometry via an octree, which is selectively pruned, enabling lossy to lossless coding. A Trisoup method can be used to refine point positions within voxels. Color attributes are encoded using either the Region-Adaptive Hierarchical Transform (RAHT) or a prediction and lifting (PredLift) transform \cite{Cao21}. V-PCC, which is designed primarily for dynamic PCs, projects 3D geometry and attributes onto 2D image sequences, with depth and texture information, which are then coded using well established and highly efficient video codecs like the High Efficiency Video Coding (HEVC) and Versatile Video Coding (VVC) standards. This approach uses conventional 2D-based codecs and does not directly code the PC geometry 3D information.  

Learning-based PCC solutions followed the success of learning-based image coding \cite{Balle18}. Initial solutions apply Convolutional Neural Networks (CNN) models to dense volumetric representations of the PC voxels \cite{Quach20,Wang21,Guarda21}, but sparse tensor representations have recently  gained traction due to their efficiency and lower computational cost \cite{Liu22,Pang22,Wang23,Lazzarotto21,Frank22}. The diversity in PC geometry data, notably regarding the point density, poses a serious challenge for learning-based PCC solutions. An early solution \cite{Guarda21} has tackled this challenge by training independent models for different PC densities. More recent solutions \cite{Wang21sparse,Quach20,Wang23} dynamically optimize the number of decoded points for a given quality metric, using side information which is determined at the PC encoder and inserted in the bitstream. 

The number of solutions for learning-based coding of the PC attributes is well below that of geometry coding. Furthermore, these solutions generally achieve compression performances for attribute coding that are below those of conventional solutions \cite{Guarda25joint}. State-of-the-art PCC solutions for geometry and color more commonly leverage projection-based approaches to code the 3D color data as 2D images, via conventional or learning-based codecs \cite{Quach20attributes,Wang22attributes}. This projection-based approach has also been adopted in the recent JPEG PCC standard, which uses a double DL-based pipeline: while the PC geometry is coded using an AE-based model, the color information is coded using the recent DL-based JPEG AI image coding standard after being projected onto 2D planes.

\section{JPEG Learning-based Point Cloud Coding Standard Review}
\label{sec3}

Due to its central role in this paper, this section reviews the JPEG Pleno Learning-based Point Cloud Coding standard (JPEG PCC) for lossy coding of static PCs, focusing on the geometry component. After a brief presentation of the overall JPEG PCC geometry codec architecture, the independent DL-based coding and SR models are presented in detail. It is important to note that only the bitstream syntax and semantics, and the decoding process are normative in the JPEG PCC standard since their normative specification is essential for interoperability; however, the JPEG PCC reference software also includes an encoder solution to be able to exercise the full codec. A description of the color-coding pipeline is out of scope in this paper. For a more detailed and complete description of JPEG PCC, please refer to \cite{JpegPCCstandard,Guarda25}.

\subsection{Overall Architecture}

This section reviews the overall architecture of the JPEG PCC geometry coding pipeline, shown in Figure \ref{fig:1_diagram}, which is structured into several key modules:

\begin{figure}
    \centering
    \includegraphics[width=\linewidth]{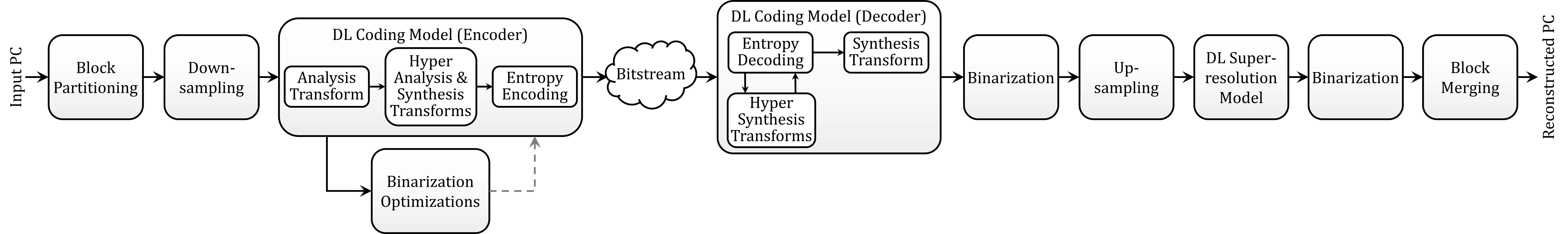}
    \caption{JPEG PCC geometry coding architecture, including the (independent) DL-based geometry coding and SR models.}
    \label{fig:1_diagram}
\end{figure}

\begin{itemize}
    \item \textbf{Block Partitioning} – The voxelized PC is divided into non-overlapping 3D blocks of a user-defined size, enabling spatial random access and a better adaptation to the available computational resources.
    \item \textbf{Down-sampling (Optional)} – Blocks may undergo resolution reduction using a SF of 1, 2 or 4, which is signaled in the bitstream (SF=1 means no down sampling is applied). This process decreases the number of points and enhances compression efficiency, particularly for sparse PCs for which down-sampling has the effect of creating denser versions of the input PC blocks. This module is also crucial to reach lower rates even for denser PCs.
    \item \textbf{DL-based Geometry Encoding} – Each PC block is encoded using a DL-based encoding model, which generates a learned latent representation for the input PC block. This model is central for this paper and will be explained in detail in the next sub-section.
    \item \textbf{Binarization Optimizations} – The output of both the coding and SR models is a set of voxel occupancy probabilities, which require a binarization strategy to reconstruct the PC voxels. A Top-k binarization strategy is adopted by JPEG PCC, in which the $k$ voxels with the higher occupancy probabilities, are selected as occupied. The value for $k$ is optimized at the encoder to obtain the best reconstruction quality at the output of the decoding model ($k_C$) and SR model ($k_S$) and inserted in the bitstream.
\end{itemize}

The PC geometry decoding process mirrors the encoding process as follows:
\begin{itemize}
    \item \textbf{DL-based Geometry Decoding} – The geometry bitstream is decoded using a DL model composed by three sub-networks, described in detail in the next sub-section.
    \item \textbf{Binarization} – The probabilities for the decoded PC block undergo Top-k binarization using the $k_C$ parameter mentioned above.
    \item \textbf{Up-sampling} – Blocks are rescaled using the SF value in the bitstream to restore the original PC resolution. This corresponds to simple coordinate scaling with no impact on the number of decoded occupied voxels.
    \item \textbf{DL-based Geometry Super-resolution (Optional)} – An optional DL-based SR model may be applied to refine the decoded geometry, if down-sampling with SF=2 or SF=4 is applied at encoder. This is the second core model addressed in this paper and will be described in the next sub-section.
    \item \textbf{Binarization} – When SR is used, a Top-k binarization is applied to the output of the SR model, using $k_S$, to determine the occupied voxels of the SR reconstructed block.
    \item \textbf{Block Merging} – Finally, the decoded blocks are reassembled to reconstruct the full PC.
\end{itemize}

\subsection{Point Cloud Geometry Coding Model}

This sub-section presents the JPEG PCC DL-based coding and decoding models. Both models use a sparse tensor representation \cite{Wang21sparse}. This means that only occupied voxels features (and their coordinates) are represented, thus reducing memory requirements and computational complexity. 
The coding model uses five sub-networks, presented in Figure \ref{fig:2_model} from the input (top) to the output (bottom). SpConv corresponds to a sparse convolution layer and GTSpConv corresponds to a generative transposed sparse convolution layer. The kernel size, number of input and output channels, and stride are shown between brackets, respectively. The PC geometry encoding process uses four of the five sub-networks in Figure \ref{fig:2_model} (not the Synthesis Transform), performing the following operations (also represented in Figure \ref{fig:1_diagram}):

\begin{figure}
    \centering
    \includegraphics[width=\linewidth]{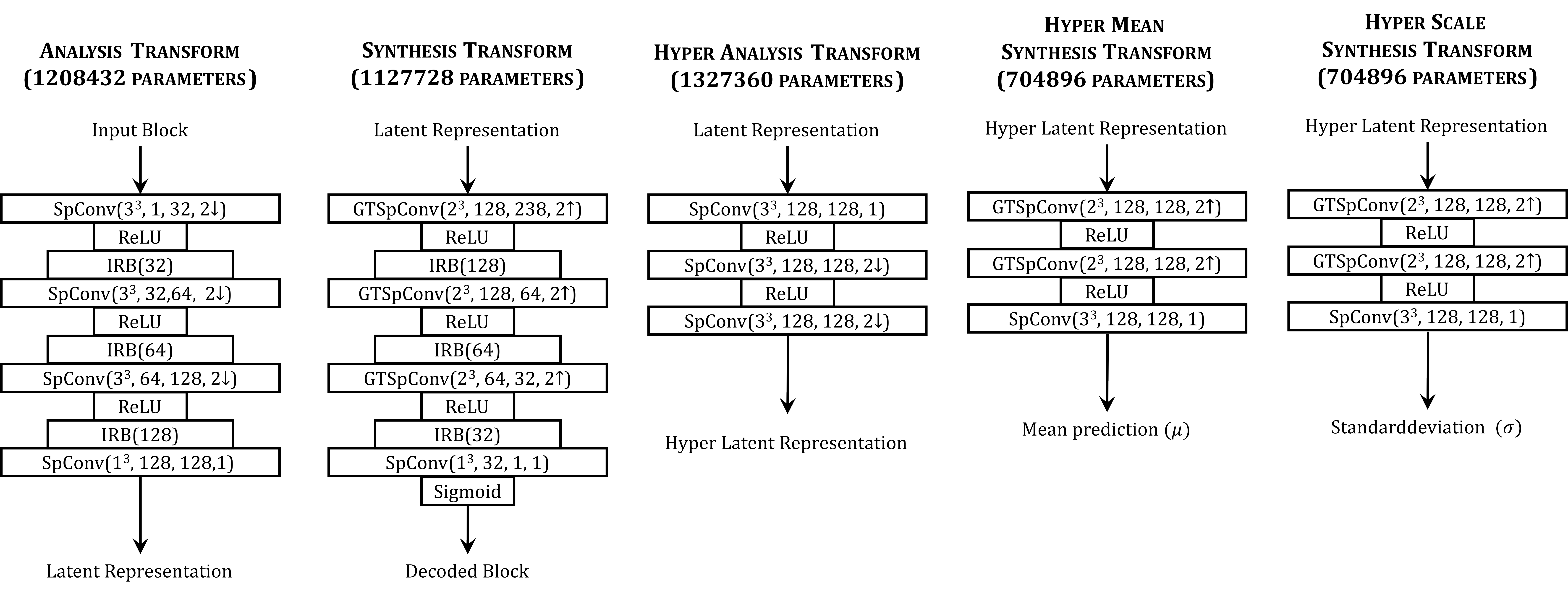}
    \caption{Detailed architecture of individual sub-networks used in the JPEG PCC geometry coding and decoding models.}
    \label{fig:2_model}
\end{figure}

\begin{itemize}
    \item \textbf{Analysis Transform} – The input PC block is converted into a latent representation using an autoencoder model (AE) composed by a series of sparse convolution layers and Inception-Residual Blocks (IRBs). While sparse convolutions reduce the spatial dimensions, IRBs extract features using several sparse convolutions with varying shape kernels, ensuring a compact yet informative representation. The latent representation is then scaled by a quantization step (QS) before entropy coding. The Analysis Transform layers progressively double the number of output channels, from 32 to 128, using an overall number of 1,208,432 parameters.
    \item \textbf{Hyper Analysis Transform} – The latent representation is processed to extract the hyper latent representation used to assist entropy coding of the latents. The Hyper Analysis Transform uses three sparse convolution layers with the same number of channels as the final layer of the Analysis Transform (128), requiring a total of 1,327,360 parameters. The hyper latents are entropy encoded and inserted in the bitstream.
    \item \textbf{Hyper Mean and Scale Synthesis Transforms} – The quantized hyper latent representation is input to two separate sub-networks, the Hyper Mean and Scale Synthesis Transforms, to generate a prediction of the means and standard deviations (i.e., scales) that characterize the probability distribution of the latent representation’s features. Both sub-networks use convolution layers with the same number of channels as the Hyper Analysis (128), requiring a total of 1.409.792 parameters, equally divided among the two.
    \item \textbf{Entropy Coding} – The latent representation’s features are entropy coded using the predicted means and scales, considering a Gaussian distribution.
    \item \textbf{Octree Encoding} – The latent and hyper latent representations’ coordinates are losslessly encoded using the G-PCC Octree encoder \cite{Cao21}. The hyper latent coordinates are derived by down-sampling the latent coordinates by a factor of four.
\end{itemize}

The decoding process receives the encoded bitstream and applies three sub-networks and an octree decoder, according to the following sequence:
\begin{itemize}
    \item \textbf{Octree Decoding} – The latent and hyper latent coordinates are decoded using the G-PCC Octree decoder.
    \item \textbf{Hyper Latent Entropy Decoder} – The bitstream is entropy decoded to generate the quantized hyper latent features which are combined with the corresponding previously decoded coordinates.
    \item \textbf{Hyper Mean and Scale Synthesis Transforms} – The hyper latents are passed to the Hyper Mean and Scale Synthesis Transforms to predict the means and scales. These are the same sub-networks used at the encoder, requiring a total of 1,409,792 parameters.
    \item \textbf{Entropy Decoding} – The latent representation’s features are entropy decoded using the predicted means and scales, considering a Gaussian distribution.
    \item \textbf{Synthesis Transform} – The reconstructed latent representation is processed by the Synthesis Transform to generate the occupancy probability of each voxel in the decoded PC block. The first layer of the Synthesis Transform model uses the same number of channels as the last layer of the Analysis Transform model (128). The following layers use a smaller number of channels in symmetric progression. The Synthesis Transform uses 1,127,728 parameters.
\end{itemize}

The DL-based geometry coding model is the backbone of JPEG PCC geometry coding pipeline which also (optionally) includes the SR model.

\subsection{Point Cloud Super-Resolution Model}

The DL-based geometry SR model is optionally used upon encoder decision and signaling but it is normatively specified when used. It has a key role in improving the JPEG PCC compression performance as it enables detail recovery from the down-sampling process applied at the encoder, at virtually no rate cost. SR uses a standalone model, which is trained independently from the coding models, and applied after decoding the PC. The SR module may be applicable if SF=2 or 4; its use is decided and signaled by the encoder, through the bitstream, when deemed advantageous in terms of compression performance.

The SR model follows a U-Net architecture \cite{unet} inspired by \cite{Akhtar20}. Like the DL geometry coding model, it operates using a more efficient sparse tensor representation. The SR model, represented in Figure \ref{fig:3_srmodel}, consists of two primary paths:

\begin{figure}
    \centering
    \includegraphics[width=\linewidth]{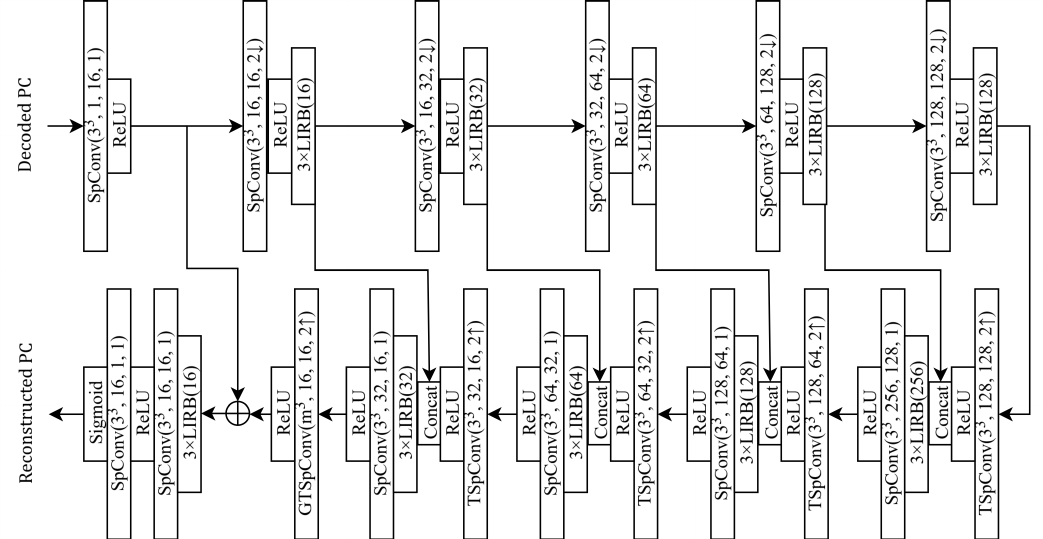}
    \caption{JPEG PCC Geometry SR model detailed architecture.}
    \label{fig:3_srmodel}
\end{figure}

\begin{itemize}
    \item \textbf{Contracting Path} – Extracts multi-scale spatial features using five sets of down-sampling sparse convolution layers followed by three Light-weight Inception-Residual Blocks (LIRBs) \cite{Guarda25joint}.
    \item \textbf{Aggregating Path} – Aggregates features from different resolution levels and progressively up samples the resolution. At each resolution level, an up-sampling transposed sparse convolution (TSpConv) increases resolution while concatenating features from the corresponding level in the contracting path. The up-sampled features then pass through three LIRBs and a final sparse convolution layer. A final up-sampling step employs a generative transposed sparse convolution (GTSpConv) to generate new points.
\end{itemize}

GTSpConv is a key distinctive feature of the JPEG PCC SR model and differs from TSpConv on the way new points are generated: while TSpConv maintains original coordinates from the contracting path, GTSpConv generates new points at all kernel positions, effectively increasing surface density. Two different SR models are available in JPEG PCC, one for SF=2 and another for SF=4:
\begin{itemize}
    \item \textit{SR model for SF=2}: Uses a kernel size ($m$) of 3 in the GTSpConv layer, resulting in a total number of 7,253,817 parameters;
    \item \textit{SR model for SF=4}: Uses a kernel size ($m$) of 5, generating more points for better surface reconstruction, resulting in a total number of 7,278,905 parameters.
\end{itemize}

\subsection{Training Process}

The training of the JPEG PCC geometry coding models (one per RD point) uses a dataset defined in the JPEG Pleno PCC Common Training and Testing Conditions (CTTC) \cite{ctc}. The training loss function uses a Rate-Distortion (RD) formulation with a Lagrangian multiplier, $\lambda$:
\begin{equation}
\label{eq1}
Loss Function = Distortion + \lambda \times Estimated Rate,
\end{equation}
where the distortion term corresponds to the average binary classification error for each voxel in the block, which is determined using the Focal Loss \cite{Lin17}. The estimated rate for each block corresponds to the joint entropies of the latents and hyper latents. To cover a wide range of RD trade-offs, five coding models were trained using different values of $\lambda$ (0.05, 0.025, 0.01, 0.005, and 0.0025); the rate associated to each of these models depends on the specific PC and coding configuration. The five models are trained sequentially: the model corresponding to the lowest $\lambda$ is trained first, using random initialization, and the following models are trained after initialization with the corresponding previous model’s weights. As a result, and since no profiles with less coding models have been defined, JPEG PCC codec requires storing the parameters associated to the five coding models; it is important to note that while the decoder parameters are normative, the same does not happen for the encoder parameters, although the JPEG PCC reference software includes five (informative) encoder models.

The training of the two SR models is done independently for the coding model, using down-sampled versions of the original (uncompressed) blocks in the JPEG PCC training dataset, for the corresponding SF=2 or SF=4. The two models are used for all target RD points, i.e., there are no models trained for specific values of $\lambda$. The training loss function consists only of a distortion term, which uses the Focal Loss metric.

\section{Proposed SR-enabled Coding Model}
\label{sec4}

This section proposes a novel lightweight SR-Enabled PC geometry Coding model (SR-EPCC), to upgrade the JPEG PCC standard by adopting a low complexity compressed domain SR approach, which allows a major reduction on the overall representation model complexity. The complexity reduction is achieved by using an SR-enabled coding model, which replaces the current JPEG PCC coding and SR models.

\subsection{Overall SR-EPCC Codec Architecture}

Figure \ref{fig:4_architecture} shows the overall architecture proposed for the JPEG SR-EPCC codec, including the SR-enabled coding model. The main architectural differences between the JPEG PCC and JPEG SR-EPCC codecs are:

\begin{figure}
    \centering
    \includegraphics[width=\linewidth]{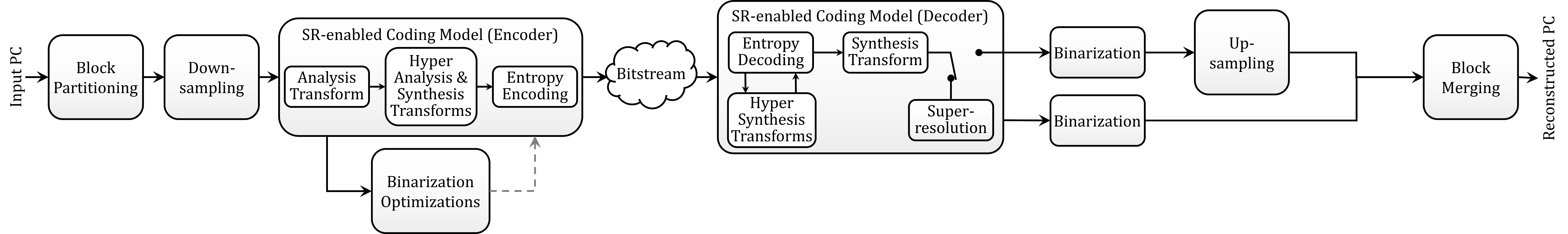}
    \caption{Proposed JPEG SR-EPCC geometry coding architecture, including the SR-enabled coding model.}
    \label{fig:4_architecture}
\end{figure}

\begin{itemize}
    \item JPEG SR-EPCC model includes a single type of DL model, i.e., the SR-EPCC model, while the JPEG PCC codec includes two types of independent models, i.e., the coding and the SR models.
    \item JPEG SR-EPCC includes a single, final voxel binarization while JPEG PCC includes two voxel binarizations, one after AE decoding and another after SR reconstruction.
    \item JPEG SR-EPCC includes an up-sampling module after binarization only when SR is not used, since up-sampling is part of the SR model when SR is used (for SF=2 and 4).
    \item For JPEG SR-EPCC, the total number of models is the same as the number of $\lambda$ values used for training, since the coding and SR components are part of the same model. This is different from JPEG PCC, which uses the same number of coding models, plus two independent SR models, one for SF=2 and another for SF=4.
\end{itemize}

The JPEG SR-EPCC model includes two novel contributions towards reduction the JPEG PCC model complexity which are presented in the next sub-sections.

\subsection{Simplified Coding Model}
\label{ssec_simp}

The JPEG PCC coding model includes five sub-networks, shown in Figure \ref{fig:2_model}, notably the Analysis and Synthesis Transforms and Hyper Analysis and Synthesis Transforms. Each PC geometry coding model targets a specific rate and includes a total of 5,073,312 model parameters; the JPEG PCC standard defines five models (one for each value of $\lambda$) to offer different RD trade-offs.

A detailed analysis of the representation power of the JPEG PCC latent representation has revealed that not all the current 128 latent channels are equally useful from the quality point of view, suggesting that some simplification is possible. Figure \ref{fig:5_qua_vs_ch} shows the average reconstructed PC geometry quality for JPEG PCC as a function of the number of decoded latent channels ($N$) using the full JPEG PCC CTTC test dataset \cite{ctc}. In this experiment, the latents are generated using the original JPEG PCC geometry encoding model, with 128 channels. However, for decoding, only the $N$ latent channels with the most energy, measured as the variance of the latents within each channel, were used. The PSNR D1 and PSNR D2 curves’ evolution as a function of $N$ shows that the PC geometry quality saturates when around $N$=20 latent channels are used, thus hinting that there is no advantage in using models with a number channels above this value; since this is the case for the JPEG PCC model, which uses 128 channels, some coding model simplification seems possible. In this context, three types of JPEG PCC coding model simplifications have been considered, notably:

\begin{figure}
    \centering
    \includegraphics[width=0.8\linewidth]{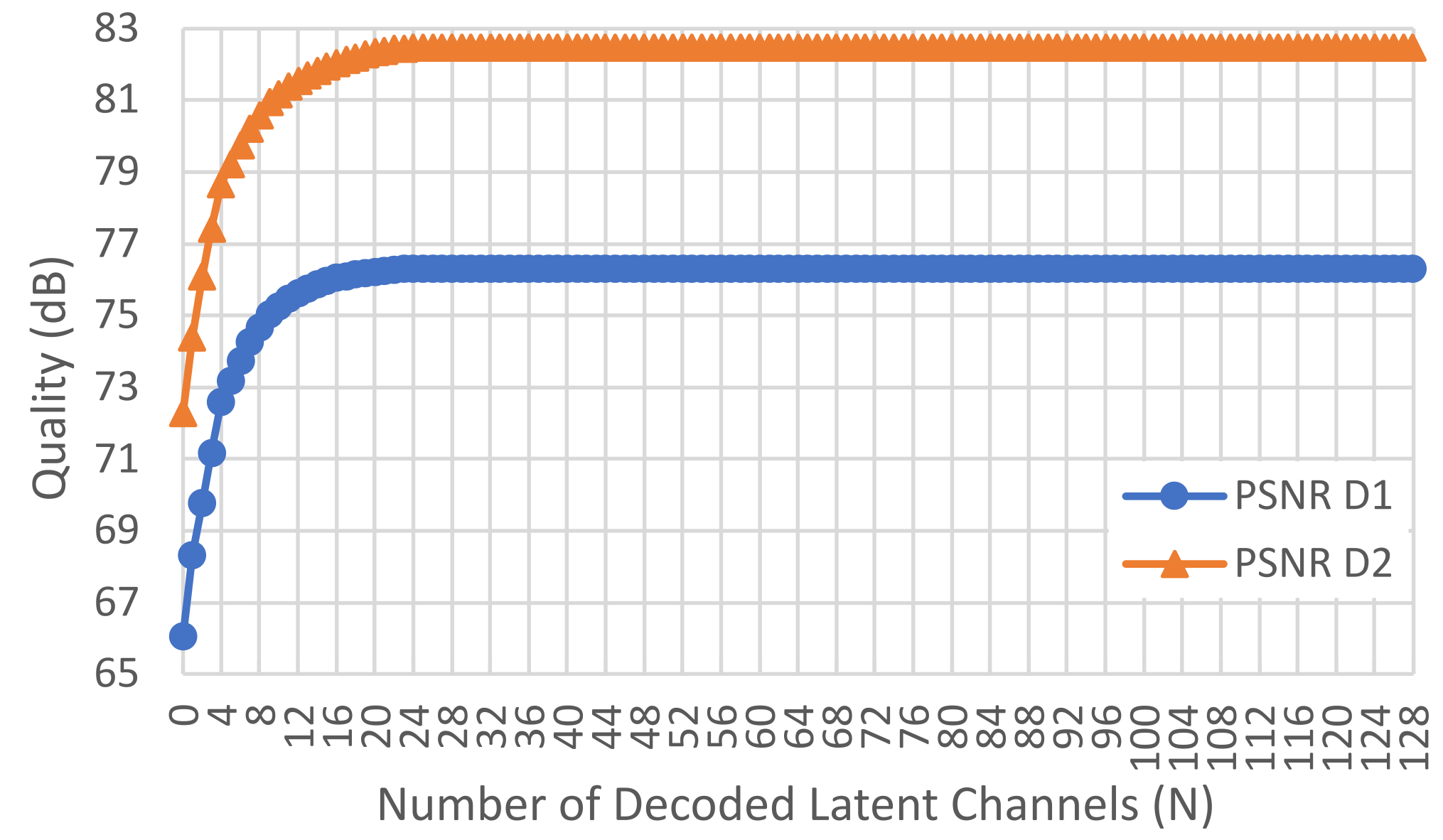}
    \caption{Average JPEG PCC reconstructed PC quality (PSNR D1 and PSNR D2) for the full JPEG PCC CTTC dataset \cite{ctc} as a function of the number of decoded latent channels, sorted in decreasing order of their variance.}
    \label{fig:5_qua_vs_ch}
\end{figure}

\begin{enumerate}
    \item \textbf{Reduce the Number of Latent Channels} – The minimal possible change to the JPEG PCC coding model implies reducing only the number of channels of the AE latent representation; this requires changing: i) the number of output channels of the last layer of the Analysis Transform and the last layer of the Mean and Scale Hyper Synthesis Transforms; and ii) the number of input channels in the first layer of the Synthesis Transform and the first layer of the Hyper Analysis Transform.
    \item \textbf{Reduce the Number of Latent and Hyper Latent Channels} – The natural next approach implies reducing the number of channels of both the latent and the hyper latent representations, since the hyper latent representation is connected to the latent representation; this requires changing: i) the number of output channels of the last layer of the Analysis Transform; ii) the number of the input channels in the first layer of the Synthesis Transform; and iii) the number of output and input channels of all layers of the Hyper Analysis and Hyper Synthesis Transforms.
    \item \textbf{Reduce All Channels} – Finally, the most drastic approach implies reducing the number of channels throughout the whole coding model; this requires changing the number of output and input channels of all layers of all sub-networks.
\end{enumerate}

These three approaches for reducing the complexity of the JPEG PCC coding model offer different trade-offs in terms of compression performance versus number of model parameters; eventual reductions on the number of model parameters depend not only on the selected simplification approach but also on the specific number of latent channels removed. 

Section \ref{ssec_results} reports results for the three model simplification approaches. The analysis of the selected simplified models’ performance is made both in terms of the achieved reduction in the number of parameters as well as the corresponding impact on compression performance. Results show that the best performance is obtained using the second approach described above, i.e., reducing the number of latent and hyper latent channels, namely reducing the number of latent and hyper latent channels from 128 to 16. The selected simplified coding model, which will be included in the proposed lightweight JPEG SR-EPCC codec, uses the novel sub-networks shown in Table \ref{tab:1_model}.

\begin{table}
    \centering
    \resizebox{\linewidth}{!}{%
    \begin{tabular}{ccccc} 
    \hline
    \textbf{Analysis Transform} & \textbf{Synthesis Transform} & \begin{tabular}[c]{@{}c@{}}\textbf{Hyper Analysis}\\\textbf{ Transform}\end{tabular} & \begin{tabular}[c]{@{}c@{}}\textbf{Hyper Mean Synthesis}\\\textbf{Transform}\end{tabular} & \begin{tabular}[c]{@{}c@{}}\textbf{Hyper Scale Synthesis}\\\textbf{Transform}\end{tabular}  \\ 
    \hline
    SpConv(3\textsuperscript{3}, 1, 32, 2↓) & GTSpConv(2\textsuperscript{3}, \textbf{16}, 128, 2↑) & SpConv(3\textsuperscript{3}, \textbf{16}, \textbf{16}, 1) & GTSpConv(2\textsuperscript{3}, \textbf{16}, \textbf{16}, 2↑) & GTSpConv(2\textsuperscript{3}, \textbf{16}, \textbf{16}, 2↑) \\
    ReLU & ReLU & ReLU & ReLU & ReLU \\
    IRB(32) & IRB(128) & SpConv(3\textsuperscript{3}, \textbf{16}, \textbf{16}, 2↓) & GTSpConv(2\textsuperscript{3}, \textbf{16}, \textbf{16}, 2↑) & GTSpConv(2\textsuperscript{3}, \textbf{16}, \textbf{16}, 2↑) \\
    SpConv(3\textsuperscript{3}, 32, 64, 2↓) & GTSpConv(2\textsuperscript{3}, 128, 64, 2↑) & ReLU & ReLU & ReLU \\
    ReLU & ReLU & SpConv(3\textsuperscript{3}, \textbf{16}, \textbf{16}, 2↓) & SpConv(3\textsuperscript{3}, \textbf{16}, \textbf{16}, 1) & SpConv(3\textsuperscript{3}, \textbf{16}, \textbf{16}, 1) \\
    IRB(64) & IRB(64) & ~ & ~ & ~ \\
    SpConv(3\textsuperscript{3}, 64, 128, 2↓) & GTSpConv(2\textsuperscript{3}, 64, 32, 2↑) & ~ & ~ & ~ \\
    ReLU & ReLU & ~ & ~ & ~ \\
    IRB(128) & IRB(32) & ~ & ~ & ~ \\
    SpConv(1\textsuperscript{3}, 128, \textbf{16}, 1) & SpConv(1\textsuperscript{3}, 32, 1, 1) & ~ & ~ & ~ \\
    \textbf{~} & Sigmoid & \textbf{~} & \textbf{~} & \textbf{~} \\ 
    \hline
    \textbf{1,194,096 parameters} & \textbf{1,013,040 parameters} & \textbf{207,68 parameters} & \textbf{11,056 parameters} & \textbf{11,056 parameters} \\
    \hline
    \end{tabular}
    }
    \caption{Detailed architecture of individual sub-networks in the simplified coding model, with changes marked in \textbf{bold}.}
    \label{tab:1_model}
\end{table}

It is relevant to notice that the selected Simplified Coding Model allows a reduction on the number of coding model parameters of about 56\% while simultaneously achieving compression gains of 1.7\% and 1.2\% for PSNR D1 and PSNR D2, respectively. The improvement in RD performance may be due to the fact that a smaller neural network (i.e., with fewer parameters) is able to be trained in a more efficient way. As a result, the smaller model is more efficient (from an RD point of view) in embedding the key features of the input PC into the latent information.

\subsection{Compressed Domain Super-Resolution Models}

The main novel contribution of this paper towards reducing the JPEG PCC model complexity is the adoption of a compressed domain SR approach to substitute the U-Net based SR solution adopted in JPEG PCC. The key idea is that SR is not applied in the spatial decompressed domain, i.e., after decoding with voxel binarization, but directly on the compressed domain, skipping the final decoding model layer and the full reconstruction of the PC data. Furthermore, by using latent information instead of the artifact-prone decoded PC data, the compressed domain SR model is able to achieve a good quality with a significant reduction in the number of model parameters.

Contrary to the standalone SR models used in JPEG PCC, which are trained with uncompressed data and do not take into account the compression artifacts or the coding rate, the compressed domain SR model is a key component of the overall SR-EPCC model, trained together with the coding model. As a result, SR-EPCC requires as many models (coding and SR) as the target rates ($\lambda$), with no additional (separate) SR models.

In the proposed SR-EPCC model architecture, the partially decoded PC latent information, produced by the first layers of the Synthesis Transform model (up to the last sparse convolution layer), is fed into a multi-branch model, as shown in Figure \ref{fig:6_cd_sr_model}. The main characteristics of the SR-EPCC geometry coding model are:

\begin{figure}
    \centering
    \includegraphics[width=0.8\linewidth]{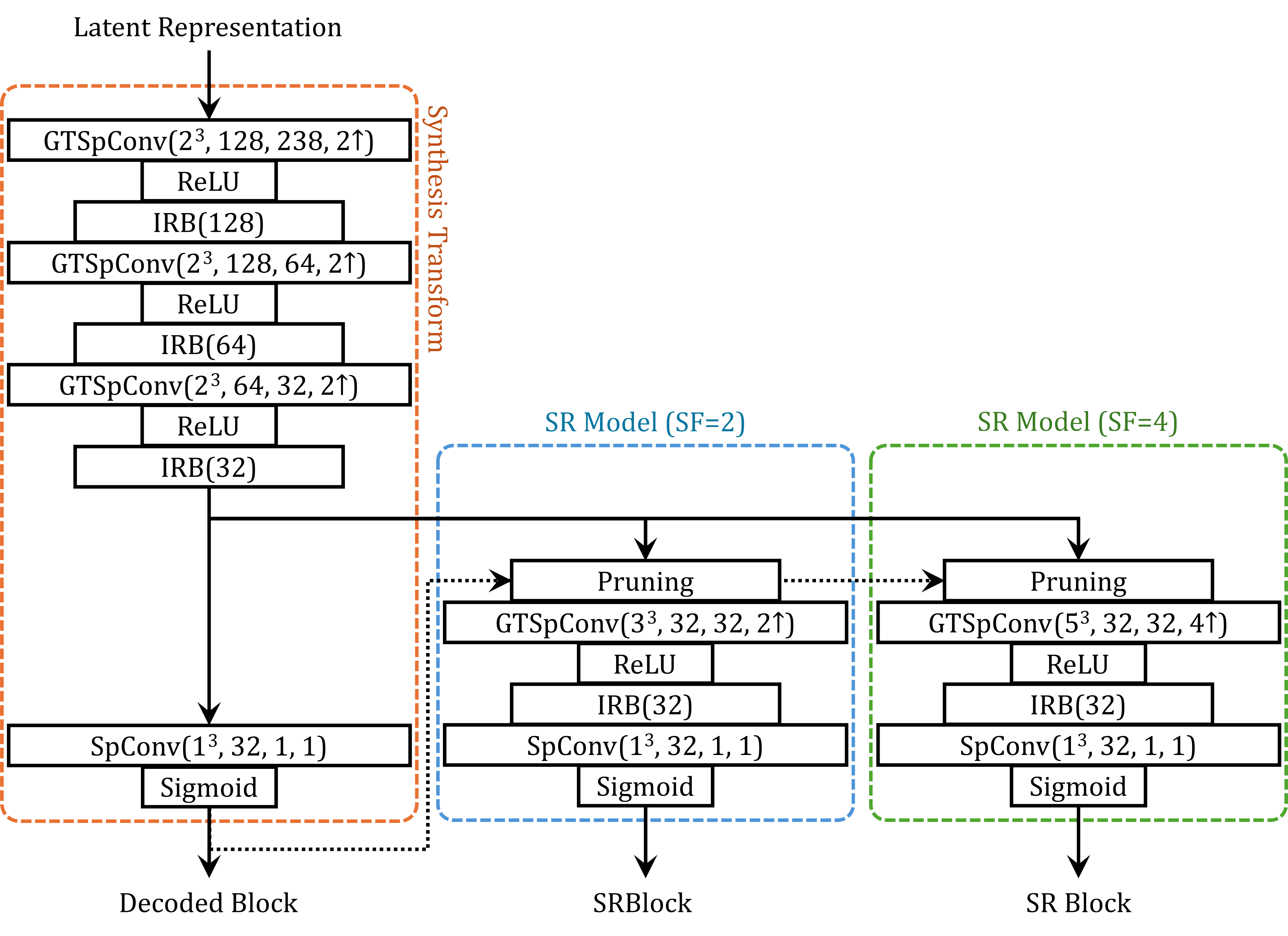}
    \caption{Detailed architecture of the Synthesis Transform and SR Models in the proposed SR-enabled Coding Model.}
    \label{fig:6_cd_sr_model}
\end{figure}

\begin{itemize}
    \item The SR-EPCC model uses three branches, corresponding to the three SR possible solutions currently supported by JPEG PCC, notably one for when SR is not used, and two for SR with SF=2 and SF=4.
    \item The SR models’ architectures are a natural extension of the Synthesis Transform architecture, which consists of three consecutive up-sampling and feature extraction layers.
    \item The SR models’ architecture uses an up-sampling generative sparse convolution layer, followed by an IRB for feature extraction and a final classification layer; the difference between SF=2 and SF=4 is in the kernel size and stride of the up-sampling layer.
    \item Prior to the SR models up-sampling layer, voxel pruning is performed in the SR branches, using the voxel occupancy probabilities produced by the classification layer (last layer) of the Synthesis Transform.
    \item In summary, the SR-EPCC model includes a coding model component corresponding to the Simplified Coding Model and a SR model component corresponding to the two SR models.
\end{itemize}

As shown in Figure \ref{fig:6_cd_sr_model}, there is the need to include a so-called \textit{pruning layer} which plays a critical role in the final RD performance. Since each up-sampling generative sparse convolution layer in the Synthesis Transform creates new non-empty voxels, the fourth layer added consecutively with the SR models would exponentially increase the number of non-empty voxels, causing poor reconstruction quality and memory issues. By pruning the voxels with a Top-k approach prior to the fourth up-sampling generative sparse convolution layer in the SR models, based on the occupancy probabilities obtained with the Synthesis Transform, the non-empty voxels are largely reduced and refined to obtain better reconstruction quality. As in the current JPEG PCC, two values for binarization optimization parameters, $k$, are determined during encoding and inserted in the bitstream: $k_C$, used to obtain the decoded block without SR and for the pruning layers of the SR models, and $k_S$, used to obtain the SR reconstructed block for either SF=2 or SF=4.

\subsection{Training Process}

As for any DL model, the training process is critical for the final performance. To accommodate the different operations performed by the novel SR-EPCC model, associated to different scales and data sizes, the training processed followed three distinct consecutive stages:
\begin{itemize}
    \item \textbf{Stage 1}: First, only the coding model layers (encoder and decoder) are trained, using the same conditions as the current JPEG PCC coding model, while ignoring all SR models’ layers. The RD loss function presented in Equation \ref{eq1} is used, and the models are trained sequentially, from the lowest to the highest values of $\lambda$ (corresponding to the highest to the lowest coding rates).
    \item \textbf{Stage 2}: Second, only the SR model layers for SF=2 are trained, while the coding model layers are frozen and the SR model layers for SF=4 are ignored. The training is performed independently for each RD trade-off ($\lambda$). In this case, since the coding model layers are frozen, the rate is constant, leading to a loss function only considering the distortion between the original and SR PCs, measured with the Focal Loss.
    \item \textbf{Stage 3}: Finally, only the SR model layers for SF=4 are trained, while freezing the coding model layers and ignoring the SR model layers for SF=2. Once again, the training is done independently for each RD trade-off ($\lambda$), and the loss function is purely the distortion between the original and SR PCs, measured with the Focal Loss.
\end{itemize}

While this multistage training strategy optimizes each model branch independently, it trains the proposed SR-EPCC model so that the impact of compression is considered when training the SR models, unlike the current JPEG PCC SR model.

\section{Performance Assessment}
\label{sec5}

This section reports the performance assessment for the proposed JPEG SR-EPCC codec in comparison with relevant benchmarks under an appropriate experimental setup.

\subsection{Experimental Setup}
This sub-section presents the experimental setup used for the performance assessment, including the datasets, benchmarks, metrics and coding configurations.

\begin{figure}
    \begin{subfigure}{.25\textwidth}
        \centering
        \includegraphics{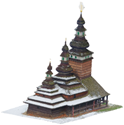}
        \caption{\textit{StMichael}}
    \end{subfigure}%
    \begin{subfigure}{.25\textwidth}
        \centering
        \includegraphics{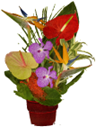}
        \caption{\textit{Bouquet}}
    \end{subfigure}%
    \begin{subfigure}{.25\textwidth}
        \centering
        \includegraphics{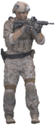}
        \caption{\textit{Soldier}}
    \end{subfigure}%
    \begin{subfigure}{.25\textwidth}
        \centering
        \includegraphics{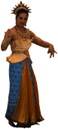}
        \caption{\textit{Thaidancer}}
    \end{subfigure}
    
    \begin{subfigure}{.2\textwidth}
        \centering
        \includegraphics{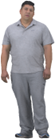}
        \caption{\textit{Boxer}}
    \end{subfigure}%
    \begin{subfigure}{.2\textwidth}
        \centering
        \includegraphics[width=\linewidth]{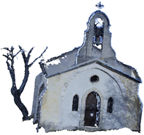}
        \caption{\textit{HouseWoRoof}}
    \end{subfigure}%
    \begin{subfigure}{.3\textwidth}
        \centering
        \includegraphics[width=\linewidth]{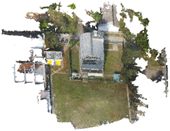}
        \caption{\textit{CITIUSP}}
    \end{subfigure}%
    \begin{subfigure}{.3\textwidth}
        \centering
        \includegraphics[width=\linewidth]{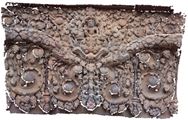}
        \caption{\textit{Facade9}}
    \end{subfigure}
    
    \begin{subfigure}{.3\textwidth}
        \centering
        \includegraphics{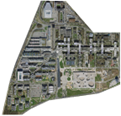}
        \caption{\textit{EPFL}}
    \end{subfigure}%
    \begin{subfigure}{.2\textwidth}
        \centering
        \includegraphics[width=\linewidth]{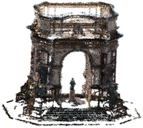}
        \caption{\textit{ArcoValentino}}
    \end{subfigure}%
    \begin{subfigure}{.2\textwidth}
        \centering
        \includegraphics[width=\linewidth]{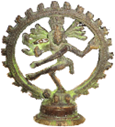}
        \caption{\textit{Shiva}}
    \end{subfigure}%
    \begin{subfigure}{.3\textwidth}
        \centering
        \includegraphics[width=\linewidth]{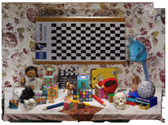}
        \caption{\textit{Unicorn}}
    \end{subfigure}
    \caption{PC test dataset for the RD performance assessment, as defined in the JPEG CTTC \cite{ctc}.}
    \label{fig:7_dataset}
\end{figure}

\textbf{Datasets} – The training and test datasets are those defined in the JPEG Pleno PCC CTTC \cite{ctc}. The test dataset, presented in Figure \ref{fig:7_dataset}, consists in 12 static PCs which were selected to ensure diversity in object and scene characteristics, particularly in terms of precision (bit depth), density and homogeneity; for a detailed statistical characterization of the test dataset, please see \cite{ctc}. Following MPEG’s classification \cite{mpegsparsity}, PCs are categorized into three groups based on density, measured by the density factor: solid, dense, and sparse. Density significantly impacts compression efficiency, making it a crucial PC classification criterion.

\textbf{Benchmarks} – To perform a solid assessment, relevant benchmarks are used, notably the (conventional) MPEG PCC standards and state-of-the-art DL-based PC geometry coding solutions with publicly available software. For this purpose, the following benchmarks, coding configurations and specific software have been used:
\begin{itemize}
    \item \textbf{G-PCC Octree v24}: MPEG G-PCC standard reference software version 24 (available in \cite{tmc13}), in the Octree geometry coding mode with specific coding configurations for each PC as specified in the JPEG Pleno PCC CTTC \cite{ctc}.
    \item \textbf{V-PCC Intra v23}: MPEG V-PCC standard reference software version 23 (available in \cite{tmc2}), with specific coding configurations for each PC as specified in the JPEG Pleno PCC CTTC \cite{ctc}.
    \item \textbf{PCGCv2 \cite{Wang21sparse}}: PCGCv2 codec (available in \cite{pcgcv2}), with the default coding configurations as specified in the software.
    \item \textbf{PCGFormer \cite{Liu22}}: PCGFormer codec (available in \cite{pcgformer}), with the default coding configurations as specified in the software, namely considering the same parameters as specified for PCGCv2.
    \item \textbf{GRASP-Net \cite{Pang22}}: GRASP-Net codec (available in \cite{graspnet}), with default coding configurations as specified in the software, namely considering the independent models trained specifically for each PC category.
\end{itemize}

\textbf{JPEG PCC-based Codecs} – Three simplified JPEG PCC-based codecs are proposed in this paper (in addition to the JPEG PCC benchmark) and assessed here, notably:
\begin{itemize}
    \item \textbf{JPEG PCC} - Corresponds to the JPEG PCC codec using the Verification Model software version 4.1 with the coding configurations for each PC as defined in \cite{Guarda25}.
    \item \textbf{JPEG SCM} – Corresponds to JPEG SR-EPCC with the proposed Simplified Coding Model (SCM) but without the compressed domain SR.
    \item \textbf{JPEG SR} – Corresponds to SR-EPCC with the proposed compressed domain SR model but without SCM.
    \item \textbf{JPEG SR-EPCC} – Corresponds to the proposed lightweight coding solution, including SCM as the coding model component and the compressed domain SR models as the SR model component.
\end{itemize}

\textbf{Metrics} – To assess the RD performance, appropriate metrics have been adopted following the JPEG Pleno PCC CTTC \cite{ctc}. The geometry rate measures the rate for the PC geometry alone and is defined as the number of bits in the geometry bitstream divided by the total number of points in the original PC, expressed in bits per point (bpp). For geometry quality assessment, two metrics are used: PSNR D1 based on a point-to-point distance; and PSNR D2 based on a point-to-plane distance \cite{ctc}. To compare the RD performance of two codecs where one is taken as reference, the Bjontegaard-Delta Rate (BD-Rate) and Bjontegaard-Delta PSNR (BD-PSNR) are used. A negative BD-Rate indicates a reduction in bitrate for the same quality, while a positive BD-PSNR reflects an improvement in quality at the same rate. Both cases imply better RD performance compared to the reference codec.

\textbf{Coding Configurations} – To ensure a fair comparison with the benchmark codecs, the JPEG Pleno PCC CTTC \cite{ctc} were followed, including the coding configurations for the G-PCC Octree and V-PCC Intra benchmarks. For JPEG PCC (and its derivations), four target rates for each PC were selected following \cite{Guarda25}, with the optimal coding configurations being determined by systematically evaluating different combinations of the available coding models (for different $\lambda$ values, providing different RD trade-offs) and coding parameters (including SF and SR options), as described in \cite{Guarda25}.

\subsection{Complexity Assessment}

This sub-section reports the complexity assessment for the JPEG PCC-based codecs in terms of the number of model parameters. Table \ref{tab:2_params} shows the number of model parameters and the corresponding reduction for the proposed SR-EPCC in comparison to JPEG PCC, when considering the coding model component and the SR model components individually, or all models together; furthermore, when applicable, it also details the number of model parameters for a single $\lambda$ value and the sum for all five $\lambda$ value. From the table, the following observations may be derived:

\begin{table}
    \centering
    \resizebox{\linewidth}{!}{%
    \begin{tabular}{c|cc|cc|cc|cc} 
    \hline
    \textbf{Component} & \multicolumn{2}{c|}{\textbf{Coding Model}} & \multicolumn{2}{c|}{\textbf{SR Model (SF=2)}} & \multicolumn{2}{c|}{\textbf{SR Model (SF=4)}} & \multicolumn{2}{c}{\textbf{All Required Models }}  \\
     & \textbf{Each $\lambda$} & \textbf{Total} & \textbf{Each $\lambda$} & \textbf{Total} & \textbf{Each $\lambda$} & \textbf{Total} & \textbf{Total} & \textbf{Reduction (\%)} \\ 
    \hline
    \textbf{JPEG PCC} & 5073312 & 25366560 & - & 7253817 & - & 7278905 & 39899282 & - \\
    \textbf{JPEG SCM} & 2250016 & 11250080 & - & 7253817 & - & 7278905 & 25782802 & 35\% \\
    \textbf{JPEG SR} & 5073312 & 25366560 & 71312 & 356560 & 171664 & 858320 & 26581440 & 33\% \\
    \textbf{JPEG SR-EPCC} & 2250016 & 11250080 & 71312 & 356560 & 171664 & 858320 & 12464960 & 69\% \\ 
    \hline
    \textbf{Reduction (\%)} & 56\% & 56\% & - & 95\% & - & 88\% & - & - \\
    \hline
    \end{tabular}
    }
    \caption{Number of model parameters in the JPEG-based PCC codecs.}
    \label{tab:2_params}
\end{table}

\begin{itemize}
    \item Compared with the original JPEG PCC geometry coding model, the coding model component of the JPEG SR-EPCC model achieves 56\% reduction on the number of parameters.
    \item Compared with the original U-Net based SR model used in JPEG PCC, the SR models/branches in the proposed compressed domain SR-EPCC model (shown in Figure \ref{fig:6_cd_sr_model}) achieve a 95\% and 88\% reduction on the number of parameters, for SF=2 and SF=4, respectively. This is even considering that JPEG SR-EPCC uses five different SR models (for each SF), trained for different RD trade-offs, whereas JPEG PCC uses only one SR model (for each SF).
    \item Considering the sum of all parameters for all required models (which vary for JPEG PCC and SR-EPCC, as previously described), the proposed JPEG SR-EPCC codec achieves a reduction on the total number of model parameters of 69\%, what is a major reduction.
\end{itemize}

The significant reduction in the total number of model parameters and memory footprint is a major contribution, since it may be essential to boost the deployment of a JPEG PCC-based codec in lower complexity terminals and consumption environments.

\subsection{RD Performance Assessment for the JPEG PCC Codecs Family}

This sub-section reports the RD performance for the three proposed lightweight JPEG-based codecs using as reference the JPEG PCC benchmark. Figure \ref{fig:8_rd_charts} and Table \ref{tab:3_rd_jpeg} show the RD performance for the JPEG-based codecs using JPEG PCC as reference. It is important to notice that comparing the three lightweight codecs, i.e., JPEG SR-EPCC, JPEG SCM and JPEG SR, effectively corresponds to performing an ablation study, effectively assessing the impact of the isolated use of the SCM and SR models proposed in this paper. The main observations regarding the RD assessment of the proposed codecs are:

\begin{figure}
    \centering
    \includegraphics[width=0.3\linewidth]{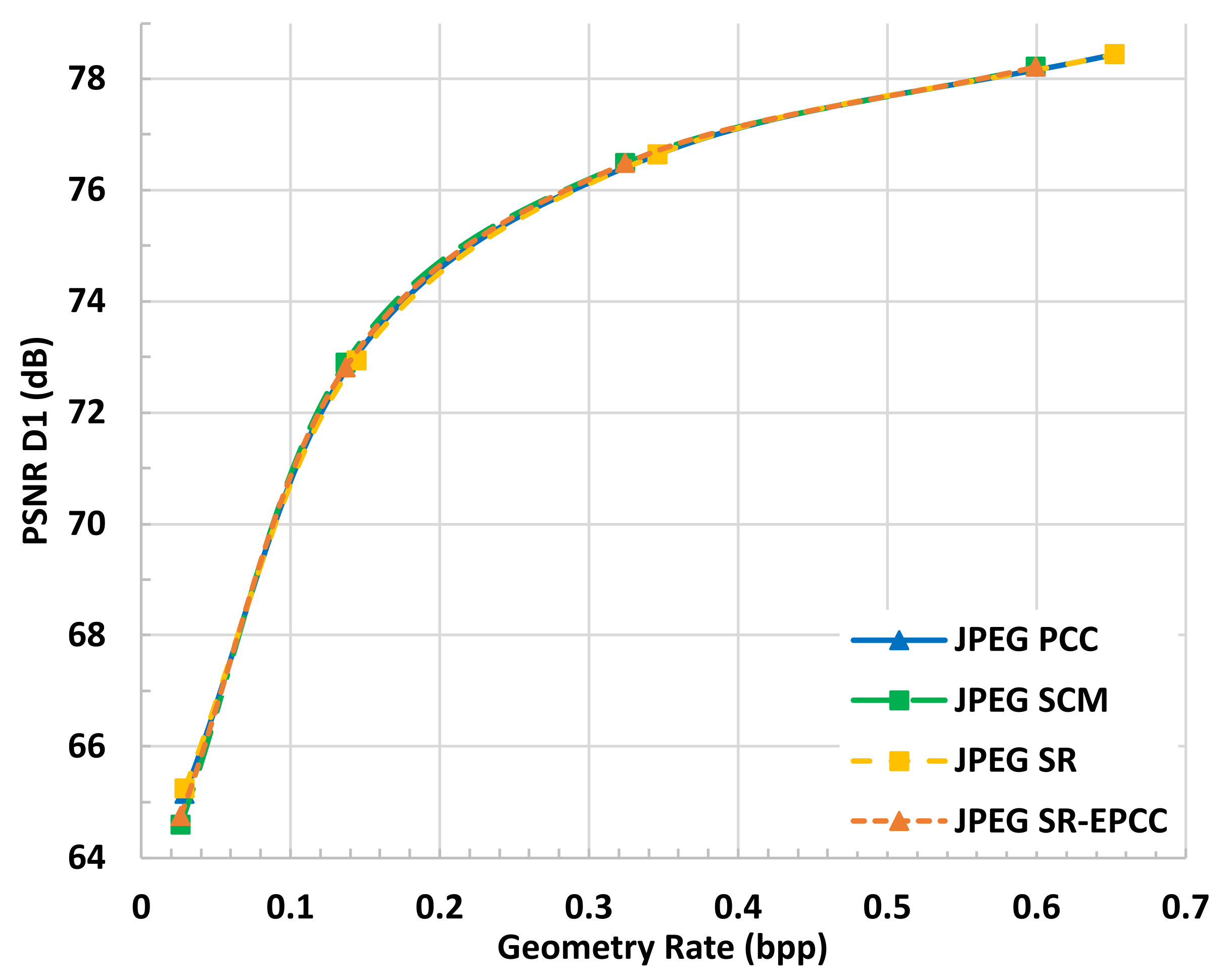}
    \includegraphics[width=0.3\linewidth]{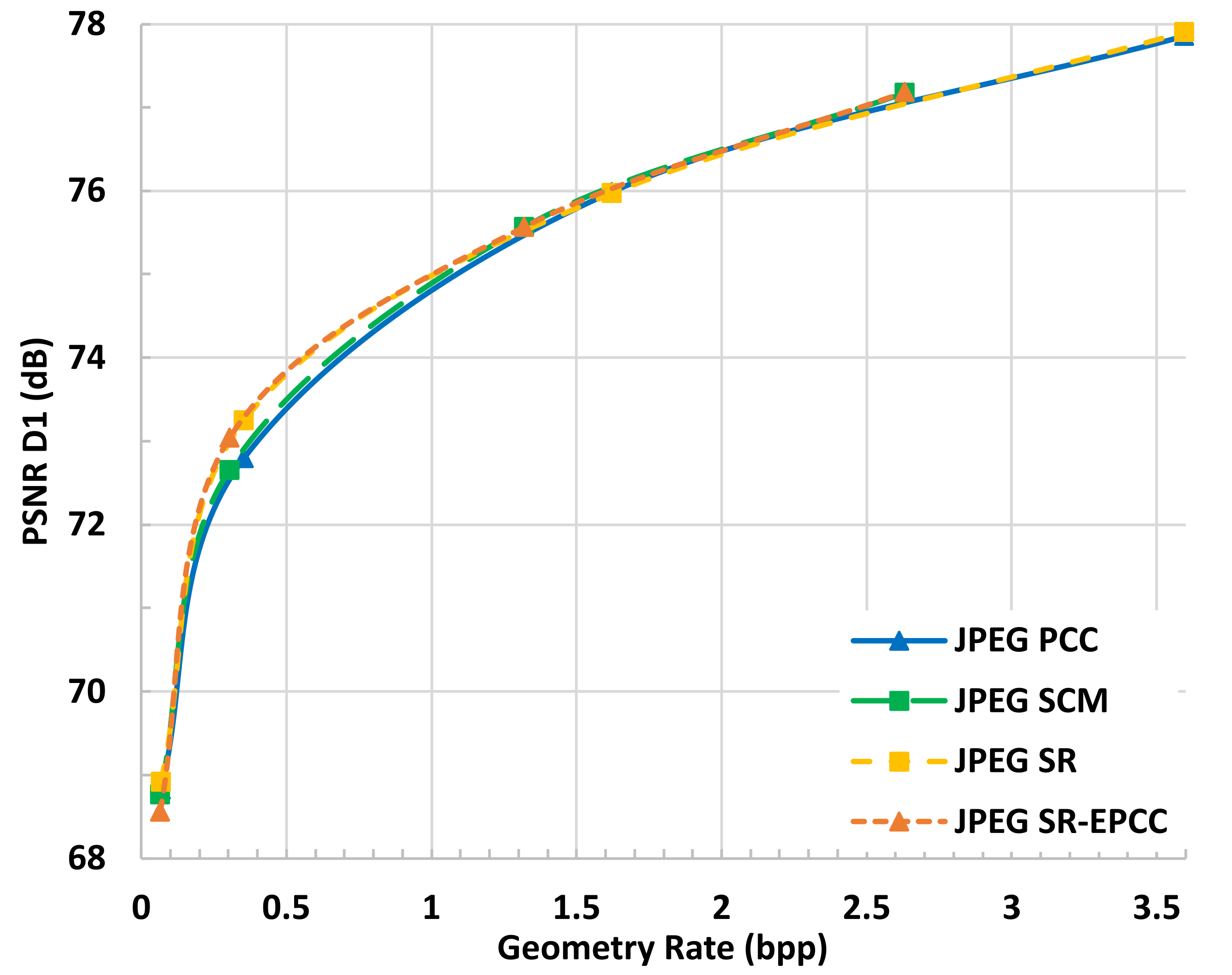}
    \includegraphics[width=0.3\linewidth]{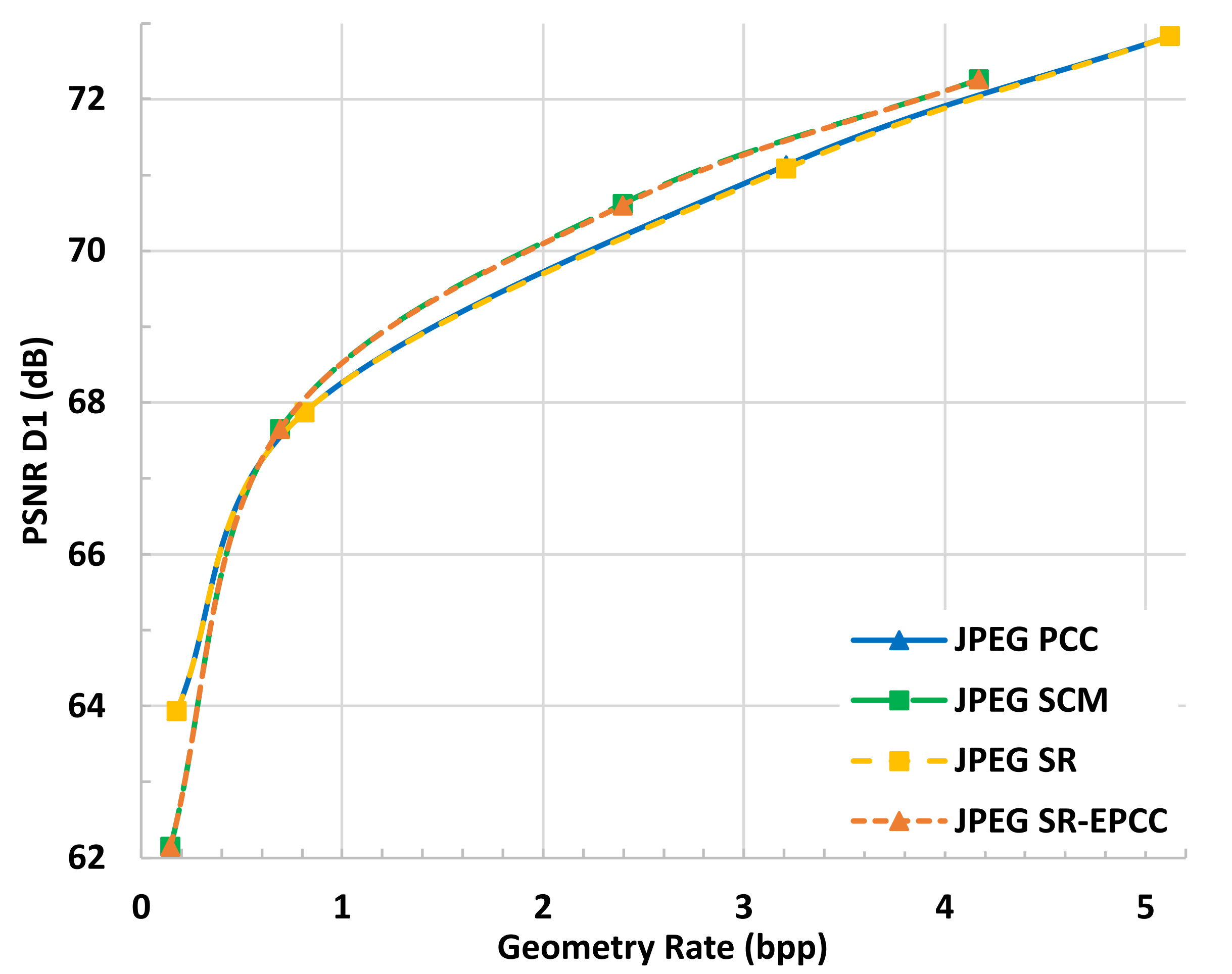}
    \includegraphics[width=0.3\linewidth]{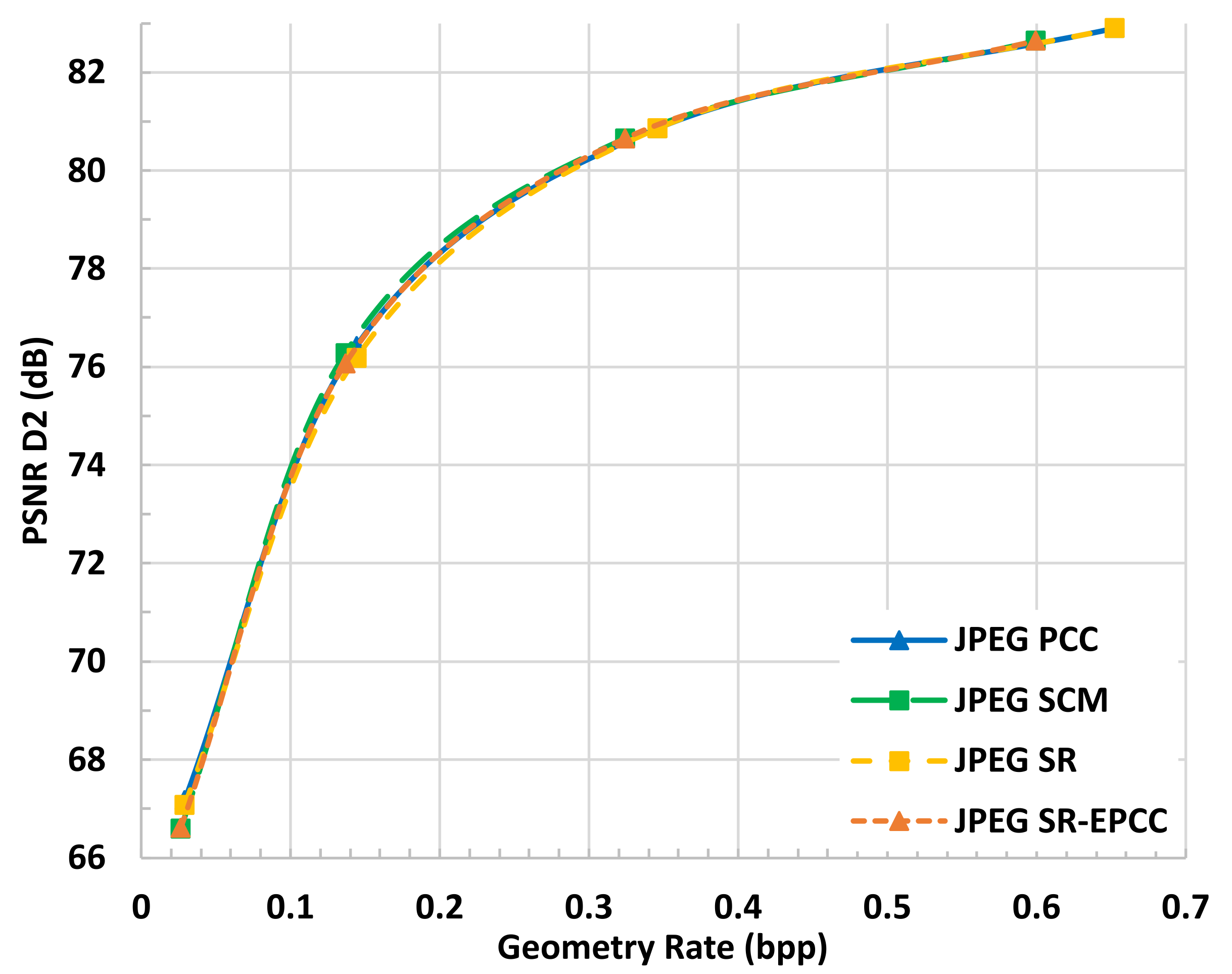}
    \includegraphics[width=0.3\linewidth]{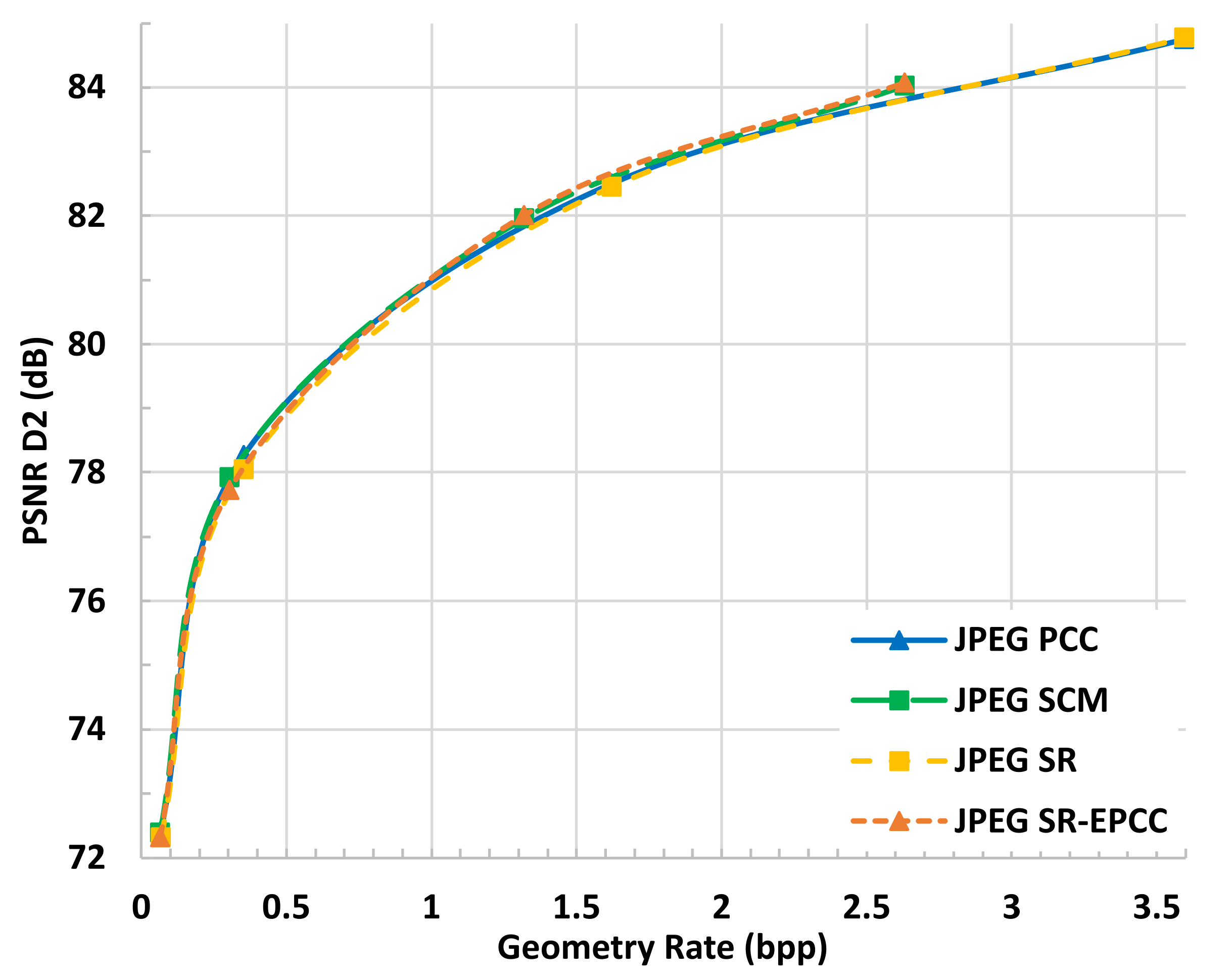}
    \includegraphics[width=0.3\linewidth]{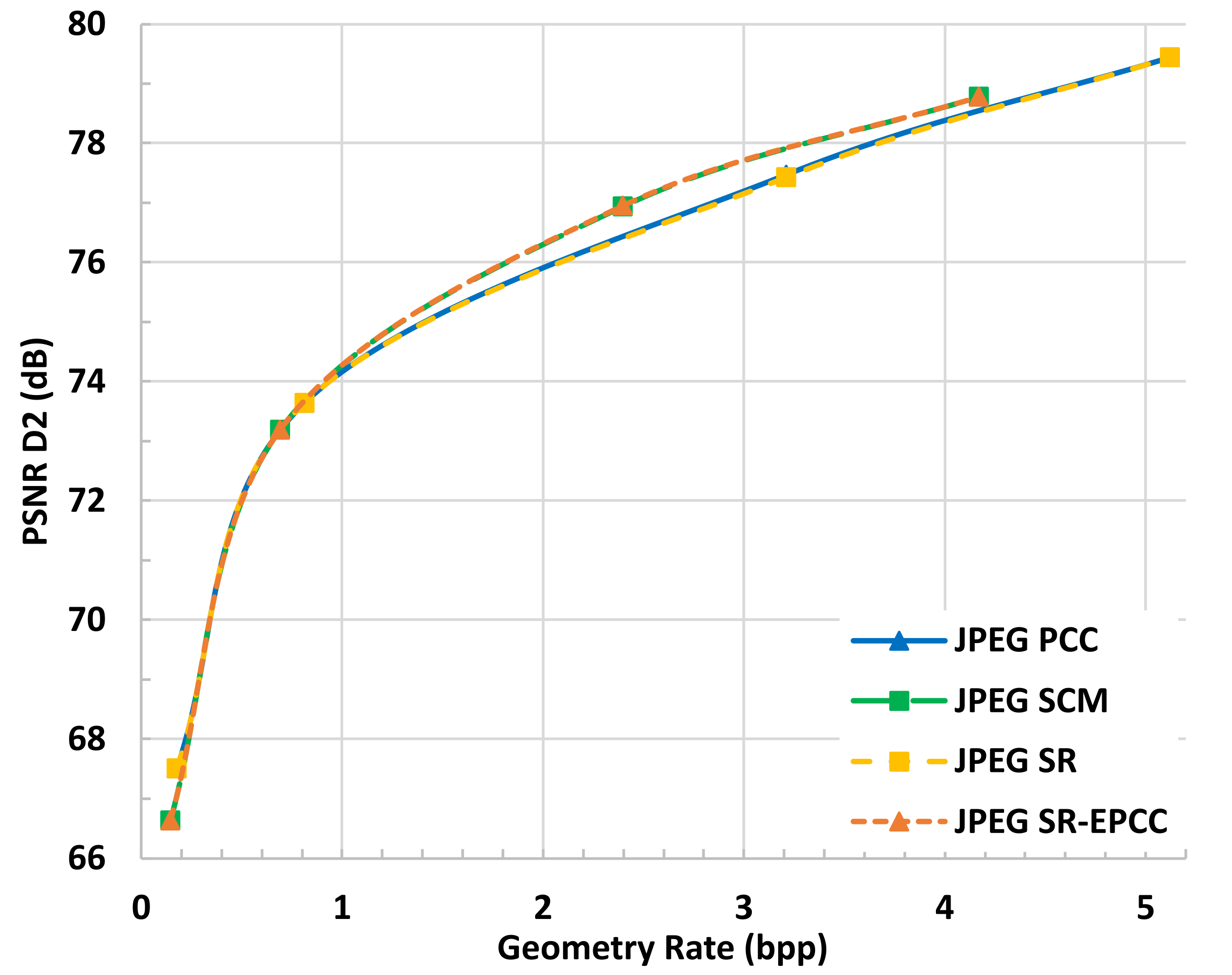}
    \caption{Average RD performance for the JPEG-based codecs: (left) Solid PCs; (middle) Dense PCs; (right) Sparse PCs.}
    \label{fig:8_rd_charts}
\end{figure}

\begin{table}
    \centering
    \resizebox{\linewidth}{!}{%
    \begin{tabular}{cl|cc|cc|cc|cc|cc|cc} 
    \hline
    \multicolumn{2}{c|}{\multirow{3}{*}{\textbf{Point Cloud}}} & \multicolumn{4}{c|}{\textbf{JPEG SCM}} & \multicolumn{4}{c|}{\textbf{JPEG SR}} & \multicolumn{4}{c}{\textbf{JPEG SR-EPCC}} \\ 
    \cline{3-14}
    \multicolumn{2}{c|}{} & \multicolumn{2}{c|}{\textbf{PSNR D1}} & \multicolumn{2}{c|}{\textbf{PSNR D2}} & \multicolumn{2}{c|}{\textbf{PSNR D1}} & \multicolumn{2}{c|}{\textbf{PSNR D2}} & \multicolumn{2}{c|}{\textbf{PSNR D1}} & \multicolumn{2}{c}{\textbf{PSNR D2}} \\ 
    \cline{3-14}
    \multicolumn{2}{c|}{} & \begin{tabular}[c]{@{}c@{}}\textbf{BD}\\\textbf{Rate}\end{tabular} & \begin{tabular}[c]{@{}c@{}}\textbf{BD}\\\textbf{PSNR}\end{tabular} & \begin{tabular}[c]{@{}c@{}}\textbf{BD}\\\textbf{Rate}\end{tabular} & \begin{tabular}[c]{@{}c@{}}\textbf{BD}\\\textbf{PSNR}\end{tabular} & \begin{tabular}[c]{@{}c@{}}\textbf{BD}\\\textbf{Rate}\end{tabular} & \begin{tabular}[c]{@{}c@{}}\textbf{BD}\\\textbf{PSNR}\end{tabular} & \begin{tabular}[c]{@{}c@{}}\textbf{BD}\\\textbf{Rate}\end{tabular} & \begin{tabular}[c]{@{}c@{}}\textbf{BD}\\\textbf{PSNR}\end{tabular} & \begin{tabular}[c]{@{}c@{}}\textbf{BD}\\\textbf{Rate}\end{tabular} & \begin{tabular}[c]{@{}c@{}}\textbf{BD}\\\textbf{PSNR}\end{tabular} & \begin{tabular}[c]{@{}c@{}}\textbf{BD}\\\textbf{Rate}\end{tabular} & \begin{tabular}[c]{@{}c@{}}\textbf{BD}\\\textbf{PSNR}\end{tabular} \\ 
    \hline
    \multirow{4}{*}{\rotatebox[origin=c]{90}{\textbf{Solid}}} & \textit{StMichael} & 2.8\% & -0.2 & 1.6\% & -0.1 & \textbf{-0.4\%} & \textbf{0.0} & 4.4\% & -0.2 & 2.4\% & -0.1 & 5.6\% & -0.3 \\
    & \textit{Bouquet} & \textbf{-7.8\%} & \textbf{0.3} & \textbf{-8.3\%}  & \textbf{0.4} & \textbf{-0.3\%} & \textbf{0.0} & 1.6\% & -0.1 & \textbf{-8.4\%}  & \textbf{0.3} & \textbf{-7.3\%}  & \textbf{0.3} \\
    & \textit{Soldier} & 0.7\% & 0.0 & 0.7\% & 0.0 & 2.1\% & -0.1 & 3.5\% & -0.2 & 1.7\% & -0.1 & 3.1\% & -0.2 \\
    & \textit{Thaidancer} & \textbf{-0.1\%} & \textbf{0.0} & \textbf{-0.2\%} & \textbf{0.0} & 2.7\% & -0.1 & 2.7\% & -0.1 & 2.5\% & -0.1 & 2.4\% & -0.1 \\ 
    \hline
    \multirow{4}{*}{\rotatebox[origin=c]{90}{\textbf{Dense}}} & \textit{Boxer} & \textbf{-4.0\%} & \textbf{0.1} & \textbf{-2.1\%} & \textbf{0.1} & 11.6\%  & -0.2 & 16.8\% & -0.5 & 5.4\% & -0.1 & 8.4\% & -0.2 \\
    & \textit{HouseWoRoof} & \textbf{-2.5\%} & \textbf{0.0} & 4.0\% & -0.1 & \textbf{-6.3\%} & \textbf{0.1} & 5.1\% & -0.1 & \textbf{-1.8\%} & \textbf{0.0} & 12.0\% & -0.3 \\
    & \textit{CITIUSP} & 0.5\% & 0.0 & \textbf{-1.7\%} & \textbf{0.1} & \textbf{-0.4\%} & \textbf{0.0} & \textbf{-0.2\%} & \textbf{0.0} & 0.4\% & 0.0 & \textbf{-2.2\%}  & \textbf{0.1} \\
    & \textit{Facade9} & \textbf{-11.5\%} & \textbf{0.3} & \textbf{-2.5\%}  & \textbf{0.1} & \textbf{-29.6\%} & \textbf{0.9} & \textbf{-1.5\%}  & \textbf{0.0} & \textbf{-34.3\%} & \textbf{1.1} & \textbf{-5.2\%} & \textbf{0.2} \\ 
    \hline
    \multirow{4}{*}{\rotatebox[origin=c]{90}{\textbf{Sparse}}} & \textit{EPFL} & \textbf{-4.7\%}  & \textbf{0.1} & \textbf{-4.3\%}  & \textbf{0.1} & 0.0\% & 0.0 & 0.0\%  & 0.0 & \textbf{-4.7\%} & \textbf{0.1} & \textbf{-4.3\%}  & \textbf{0.1} \\
    & \textit{ArcoValentino} & 0.5\% & 0.0 & \textbf{-0.7\%}  & \textbf{0.0} & 0.0\% & 0.0 & 0.0\% & 0.0 & 0.5\%   & 0.0 & \textbf{-0.7\%}  & \textbf{0.0} \\
    & \textit{Shiva} & \textbf{-4.8\%}  & \textbf{0.1} & \textbf{-5.6\%}  & \textbf{0.2} & 1.6\% & 0.0 & 1.3\%   & 0.0 & \textbf{-4.1\%} & \textbf{0.1} & \textbf{-6.0\%} & \textbf{0.2} \\
    & \textit{Unicorn} & 10.5\% & -0.5 & 4.9\% & -0.1 & 0.0\% & 0.0 & 0.0\% & 0.0 & 10.5\% & -0.5 & 4.9\% & -0.1 \\ 
    \hline
    ~ & Average Solid & \textbf{-1.1\%}  & \textbf{0.0} & \textbf{-1.5\%}  & \textbf{0.0} & 1.0\% & 0.0 & 3.1\%   & -0.1 & \textbf{-0.5\%}  & \textbf{0.0} & 1.0\% & -0.1 \\
    ~ & Average Dense & \textbf{-4.4\%}  & \textbf{0.1} & \textbf{-0.6\%}  & \textbf{0.0} & \textbf{-6.2\%} & \textbf{0.2} & 5.0\% & -0.1 & \textbf{-7.6\%}  & \textbf{0.3} & 3.3\% & -0.1 \\
    ~ & Average Sparse & 0.4\%  & -0.1 & \textbf{-1.4\%}  & \textbf{0.1} & 0.4\% & 0.0 & 0.3\% & 0.0 & 0.6\% & -0.1 & \textbf{-1.5\%}  & \textbf{0.1} \\ 
    \hline
    ~ & Average All & \textbf{-1.7\%}  & \textbf{0.0} & \textbf{-1.2\%}  & \textbf{0.0} & \textbf{-1.6\%}  & \textbf{0.1} & 2.8\% & -0.1 & \textbf{-2.5\%}  & \textbf{0.1} & 0.9\% & 0.0 \\
    \hline
    \end{tabular}
    }
    \caption{BD-Rate and BD-PSNR for the JPEG-based codecs using JPEG PCC as reference.}
    \label{tab:3_rd_jpeg}
\end{table}

\begin{itemize}
    \item Overall, the RD performance for the three lightweight codecs is similar to JPEG PCC, with some small positive or negative variations for specific PCs. The overall dataset average results even show that the already reported complexity gains coincide with a small average compression performance gain for PSNR D1. These gains are more evident for the dense PCs.
    \item The RD performance impact is slightly worse for PSNR D2, because the Top-k binarizations were optimized for the PSNR D1 quality metric; this may be changed by adopting another quality metric for the Top-k binarizations.
    \item JPEG SCM shows a similar RD performance (or even some light gains for PSNR D1) regarding JPEG PCC, thus highlighting the benefit of the Simplified Coding Model (i.e., reducing number of latent channels from 128 to 16);
    \item JPEG SR shows a similar RD performance to JPEG PCC (or even some light gains for PSNR D1), thus confirming that the compressed domain SR approach reaches the same compression at much less complexity.
    \item JPEG SR-EPCC shows the best average RD performance of the proposed lightweight solutions, when PSNR D1 is considered, with gains of 2.5\% in BD-Rate. The impact varies across PC, between a penalty in BD-Rate of 10.5\% for sparse PC \textit{Unicorn} and a gain of 34.3\% for the dense PC \textit{Facade}.
\end{itemize}

\subsection{RD Performance Assessment Against Benchmarks}
\label{ssec_results}

This sub-section reports the RD performance for JPEG SR-EPCC in comparison to the benchmarks, summarized in Table \ref{tab:4_rd_bench}. The main observations are:

\begin{table}
    \centering
    \resizebox{\linewidth}{!}{%
    \begin{tabular}{ll|cc|cc|cc|cc|cc|cc|cc|cc|cc|cc} 
    \hline
    \multicolumn{2}{c|}{\multirow{3}{*}{\textbf{Point Cloud}}} & \multicolumn{4}{c|}{\textbf{Ref: G-PCC Octree v24}} & \multicolumn{4}{c|}{\textbf{Ref: V-PCC Intra v23}} & \multicolumn{4}{c|}{\textbf{Ref: PCGCv2}} & \multicolumn{4}{c|}{\textbf{Ref: PCGFormer}} & \multicolumn{4}{c}{\textbf{Ref: GRASP-Net}} \\ 
    \cline{3-22}
    \multicolumn{2}{c|}{} & \multicolumn{2}{c|}{\textbf{PSNR D1}} & \multicolumn{2}{c|}{\textbf{PSNR D2}} & \multicolumn{2}{c|}{\textbf{PSNR D1}} & \multicolumn{2}{c|}{\textbf{PSNR D2}} & \multicolumn{2}{c|}{\textbf{PSNR D1}} & \multicolumn{2}{c|}{\textbf{PSNR D2}} & \multicolumn{2}{c|}{\textbf{PSNR D1}} & \multicolumn{2}{c|}{\textbf{PSNR D2}} & \multicolumn{2}{c|}{\textbf{PSNR D1}} & \multicolumn{2}{c}{\textbf{PSNR D2}} \\ 
    \cline{3-22}
    \multicolumn{2}{c|}{} & \begin{tabular}[c]{@{}c@{}}\textbf{BD}\\\textbf{Rate}\end{tabular} & \begin{tabular}[c]{@{}c@{}}\textbf{BD}\\\textbf{PSNR}\end{tabular} & \begin{tabular}[c]{@{}c@{}}\textbf{BD}\\\textbf{Rate}\end{tabular} & \begin{tabular}[c]{@{}c@{}}\textbf{BD}\\\textbf{PSNR}\end{tabular} & \begin{tabular}[c]{@{}c@{}}\textbf{BD}\\\textbf{Rate}\end{tabular} & \begin{tabular}[c]{@{}c@{}}\textbf{BD}\\\textbf{PSNR}\end{tabular} & \begin{tabular}[c]{@{}c@{}}\textbf{BD}\\\textbf{Rate}\end{tabular} & \begin{tabular}[c]{@{}c@{}}\textbf{BD}\\\textbf{PSNR}\end{tabular} & \begin{tabular}[c]{@{}c@{}}\textbf{BD}\\\textbf{Rate}\end{tabular} & \begin{tabular}[c]{@{}c@{}}\textbf{BD}\\\textbf{PSNR}\end{tabular} & \begin{tabular}[c]{@{}c@{}}\textbf{BD}\\\textbf{Rate}\end{tabular} & \begin{tabular}[c]{@{}c@{}}\textbf{BD}\\\textbf{PSNR}\end{tabular} & \begin{tabular}[c]{@{}c@{}}\textbf{BD}\\\textbf{Rate}\end{tabular} & \begin{tabular}[c]{@{}c@{}}\textbf{BD}\\\textbf{PSNR}\end{tabular} & \begin{tabular}[c]{@{}c@{}}\textbf{BD}\\\textbf{Rate}\end{tabular} & \begin{tabular}[c]{@{}c@{}}\textbf{BD}\\\textbf{PSNR}\end{tabular} & \begin{tabular}[c]{@{}c@{}}\textbf{BD}\\\textbf{Rate}\end{tabular} & \begin{tabular}[c]{@{}c@{}}\textbf{BD}\\\textbf{PSNR}\end{tabular} & \begin{tabular}[c]{@{}c@{}}\textbf{BD}\\\textbf{Rate}\end{tabular} & \begin{tabular}[c]{@{}c@{}}\textbf{BD}\\\textbf{PSNR}\end{tabular}  \\ 
    \hline
    \multirow{4}{*}{\rotatebox[origin=c]{90}{\textbf{Solid}}}  & \textit{StMichael} & \textbf{-84\%} & \textbf{8.2} & \textbf{-73\%} & \textbf{6.6} & \textbf{-59\%} & \textbf{4.0} & \textbf{-55\%} & \textbf{4.5} & \textbf{-14\%} & \textbf{0.6} & \textbf{-9\%} & \textbf{0.5} & 8\% & -0.1 & 5\% & \textbf{0.1} & \textbf{-22\%} & \textbf{1.1} & \textbf{-24\%} & \textbf{1.4} \\
    & \textit{Bouquet} & \textbf{-84\%} & \textbf{7.5} & \textbf{-78\%} & \textbf{6.3} & \textbf{-59\%} & \textbf{3.8} & \textbf{-59\%} & \textbf{4.2} & 13\% & -0.8 & 9\% & -0.8 & 28\% & -1.0 & 20\% & -1.0 & \textbf{-9\%} & \textbf{0.4} & \textbf{-16\%} & \textbf{0.7} \\
    & \textit{Soldier} & \textbf{-88\%} & \textbf{10.2}                                                       & \textbf{-81\%}                                                      & \textbf{9.0} & \textbf{-43\%} & \textbf{3.0} & \textbf{-42\%} & \textbf{3.5} & \textbf{-15\%} & \textbf{0.8} & \textbf{-13\%} & \textbf{0.7} & 3\% & 0.0 & 4\% & 0.0 & \textbf{-44\%} & \textbf{2.4} & \textbf{-43\%} & \textbf{3.0} \\
    & \textit{Thaidancer} & \textbf{-91\%} & \textbf{9.9} & \textbf{-83\%} & \textbf{8.9} & \textbf{-36\%} & \textbf{2.3} & \textbf{-31\%} & \textbf{2.5} & \textbf{-59\%} & \textbf{3.6} & \textbf{-16\%} & \textbf{0.9} & \textbf{-67\%} & \textbf{3.4} & 12\% & -0.2 & \textbf{-49\%} & \textbf{2.8} & \textbf{-49\%} & \textbf{3.6} \\ 
    \hline
    \multirow{4}{*}{\rotatebox[origin=c]{90}{\textbf{Dense}}}  & \textit{Boxer} & \textbf{-83\%} & \textbf{6.0} & \textbf{-88\%} & \textbf{8.5} & \textbf{-36\%} & \textbf{0.9} & \textbf{-33\%} & \textbf{1.6} & \textbf{-79\%} & \textbf{3.1} & \textbf{-63\%} & \textbf{3.9} & \textbf{-91\%} & \textbf{3.1} & \textbf{-27\%} & \textbf{2.8} & \textbf{-41\%} & \textbf{0.5} & \textbf{-40\%} & \textbf{1.3} \\
    & \textit{HouseWoRoof} & \textbf{-63\%} & \textbf{3.5} & \textbf{-58\%} & \textbf{3.7} & \textbf{-27\%} & \textbf{0.6} & \textbf{-52\%} & \textbf{2.2} & \textbf{-52\%} & \textbf{1.5} & \textbf{-64\%} & \textbf{1.6} & \textbf{-72\%} & \textbf{2.2} & \textbf{-56\%} & \textbf{1.4} & \textbf{-11\%} & \textbf{0.2} & \textbf{-26\%} & \textbf{0.7} \\
    & \textit{CITIUSP} & \textbf{-33\%} & \textbf{1.7} & \textbf{-21\%} & \textbf{1.0} & 2\% & -0.2 & 2\% & -0.2 & \textbf{-60\%} & \textbf{3.1} & 22\% & -0.3 & \textbf{-58\%} & \textbf{3.7} & 29\% & -0.1 & 21\% & -0.7 & 20\% & -0.8 \\
    & \textit{Facade9} & \textbf{-58\%} & \textbf{2.9} & \textbf{-59\%} & \textbf{3.8} & 392\% & -4.6 & 102\% & -3.2 & \textbf{-60\%} & \textbf{3.0} & \textbf{-48\%} & \textbf{2.0} & \textbf{-71\%} & \textbf{3.5} & \textbf{-40\%} & \textbf{1.8} & \textbf{-27\%} & \textbf{0.7} & \textbf{-37\%} & \textbf{1.3} \\ 
    \hline
    \multirow{4}{*}{\rotatebox[origin=c]{90}{\textbf{Sparse}}} & \textit{EPFL} & \textbf{-34\%} & \textbf{1.4} & \textbf{-19\%} & \textbf{0.7} & \textbf{-29\%} & \textbf{0.8} & \textbf{-19\%} & \textbf{0.7} & \textbf{-7\%} & \textbf{0.7} & \textbf{-16\%} & \textbf{1.3} & 10\% & \textbf{1.0} & \textbf{-38\%} & \textbf{1.4} & 5\% & -0.2 & \textbf{-16\%} & \textbf{0.2} \\
    & \textit{ArcoValentino}  & \textbf{-34\%} & \textbf{1.4} & \textbf{-24\%} & \textbf{0.8} & \textbf{-69\%} & \textbf{2.9} & \textbf{-77\%} & \textbf{4.0} & \textbf{-57\%} & \textbf{2.8} & \textbf{-50\%} & \textbf{2.6} & \textbf{-61\%} & \textbf{2.9} & \textbf{-55\%} & \textbf{2.7} & \textbf{-8\%} & -0.1 & \textbf{-8\%} & -0.1 \\
    & \textit{Shiva} & \textbf{-44\%} & \textbf{2.1} & \textbf{-35\%} & \textbf{1.7} & \textbf{-73\%} & \textbf{2.9} & \textbf{-86\%} & \textbf{5.4} & \textbf{-22\%} & \textbf{0.7} & \textbf{-12\%} & \textbf{0.8} & \textbf{-22\%} & \textbf{0.7} & 21\% & -0.3 & \textbf{-11\%} & \textbf{0.1} & \textbf{-17\%} & \textbf{0.6} \\
    & \textit{Unicorn} & 97\% & -2.3 & \textbf{-32\%} & \textbf{1.6} & 30\% & -1.0 & \textbf{-25\%} & -1.5 & \textbf{-92\%} & \textbf{17.6} & \textbf{-90\%} & \textbf{15.7} & \textbf{-88\%} & \textbf{8.8} & \textbf{-84\%} & \textbf{4.6} & 104\% & -1.8 & 19\% & \textbf{0.1} \\ 
    \hline
    ~ & Average Solid & \textbf{-87\%} & \textbf{9.0} & \textbf{-79\%} & \textbf{7.7} & \textbf{-49\%} & \textbf{3.3} & \textbf{-47\%} & \textbf{3.7} & \textbf{-19\%} & \textbf{1.0} & \textbf{-7\%} & \textbf{0.3} & \textbf{-7\%} & \textbf{0.6} & 10\% & -0.3 & \textbf{-31\%} & \textbf{1.7} & \textbf{-33\%} & \textbf{2.2} \\
    ~ & Average Dense & \textbf{-59\%} & \textbf{3.5} & \textbf{-57\%} & \textbf{4.2} & 83\% & -0.8 & 5\% & \textbf{0.1} & \textbf{-63\%} & \textbf{2.7} & \textbf{-38\%} & \textbf{1.8} & \textbf{-73\%} & \textbf{3.1} & \textbf{-23\%} & \textbf{1.5} & \textbf{-15\%} & \textbf{0.2} & \textbf{-21\%} & \textbf{0.6} \\
    ~ & Average Sparse & \textbf{-4\%} & \textbf{0.7} & \textbf{-27\%} & \textbf{1.2} & \textbf{-35\%} & \textbf{1.4} & \textbf{-52\%} & \textbf{2.2} & \textbf{-44\%} & \textbf{5.4} & \textbf{-42\%} & \textbf{5.1} & \textbf{-40\%} & \textbf{3.3} & \textbf{-39\%} & \textbf{2.1} & 23\% & -0.5 & \textbf{-6\%} & \textbf{0.2} \\ 
    \hline
    ~ & Average All & \textbf{-50\%} & \textbf{4.4} & \textbf{-54\%} & \textbf{4.4} & \textbf{-1\%} & \textbf{1.3} & \textbf{-31\%} & \textbf{2.0} & \textbf{-42\%} & \textbf{3.0} & \textbf{-29\%} & \textbf{2.4} & \textbf{-40\%} & \textbf{2.4} & \textbf{-17\%} & \textbf{1.1} & \textbf{-8\%} & \textbf{0.5} & \textbf{-20\%} & \textbf{1.0} \\
    \hline
    \end{tabular}
    }
    \caption{BD-Rate and BD-PSNR for the proposed JPEG SR-EPCC codec using each benchmark as reference.}
    \label{tab:4_rd_bench}
\end{table}

\begin{itemize}
    \item JPEG SR-EPCC performs better than all conventional and DL-based benchmarks for the average of all solid PCs, with the single exception of PCGFormer for PSNR D2. The highest BD-Rate gain is 87\% over G-PCC for PSNR D1.
    \item JPEG SR-EPCC performs better than all conventional and DL-based benchmarks for the average of all dense PCs, with the exception of V-PCC Intra for PSNR D1 and PSNR D2, although mostly due to the singular loss for the \textit{Facade} PC.
    \item JPEG SR-EPCC performs better than all conventional and DL-based benchmarks for the average of all sparse PCs with the single exception of GRASP-Net for PSNR D1.
\end{itemize}

Overall, JPEG SR-EPCC offers an exceptional RD performance regarding both the conventional and DL-based benchmarks since it is able to consistently achieve top performance for all types of PCs without applying any specific tuning to any type of PC.

\subsection{Simplified PC Geometry Coding Models Study}

This sub-section reports the results for the three model simplification approaches proposed in Section \ref{ssec_simp} to reduce the number of model parameters of the original JPEG PCC geometry coding model (corresponding to the JPEG SCM codec). Table \ref{tab:5_rd_simp} reports the BD-Rate and BD-PSNR results, together with the total number of model parameters and the corresponding percentual reduction, when different numbers of latent channels are used, for the full JPEG PCC CTTC dataset \cite{ctc}. The main observations are:

\begin{table}
    \centering
    \resizebox{\linewidth}{!}{%
    \begin{tabular}{c|c|cc|cc|cc} 
    \hline
    \multirow{2}{*}{\textbf{Simplification Approach}} & \multirow{2}{*}{\begin{tabular}[c]{@{}c@{}}\textbf{No. of} \\\textbf{Channels}\end{tabular}} & \multicolumn{2}{c|}{\textbf{PSNR D1}} & \multicolumn{2}{c|}{\textbf{PSNR D2}} & \multicolumn{2}{c}{\textbf{Model Parameters}}  \\
    & & \begin{tabular}[c]{@{}c@{}}\textbf{BD}\\\textbf{Rate}\end{tabular} & \begin{tabular}[c]{@{}c@{}}\textbf{BD}\\\textbf{PSNR}\end{tabular} & \begin{tabular}[c]{@{}c@{}}\textbf{BD}\\\textbf{Rate}\end{tabular} & \begin{tabular}[c]{@{}c@{}}\textbf{BD}\\\textbf{PSNR}\end{tabular} & \textbf{Each $\lambda$}  & \textbf{Reduction}          \\ 
    \hline
    \multirow{4}{*}{\textbf{Latent Channels}} & 64 & 1.7\% & -0.1 & 1.8\% & -0.1 & 4335904 & 15\% \\
    & 32 & 1.8\% & -0.1 & 1.9\% & -0.1 & 3967200 & 22\% \\
    & 24 & 0.2\% & 0.0 & 0.8\% & 0.0 & 3875024 & 24\% \\
    & 16 & 0.5\% & 0.0 & 0.0\% & 0.0 & 3782848 & 25\% \\ 
    \hline
    \multirow{6}{*}{\textbf{Latent and Hyper Latent Channels}} & 64 & 1.9\% & -0.1 & 1.4\% & -0.1 & 2946976 & 42\% \\
    & 32 & -0.5\% & 0.0 & 0.5\% & 0.0 & 2396832 & 53\% \\
    & 24 & -1.5\% & 0.0 & -0.9\% & 0.0 & 2312736 & 54\% \\
    & \textbf{16} & \textbf{-1.7\%} & \textbf{0.0} & \textbf{-1.2\%} & \textbf{0.0} & \textbf{2250016} & \textbf{56\%} \\
    & 12 & -1.5\% & 0.0 & 0.1\% & 0.0 & 2226672 & 56\% \\
    & 8 & -1.0\% & 0.0 & 1.2\% & 0.0 & 2208672 & 56\% \\ 
    \hline
    \multirow{4}{*}{\textbf{All Channels}} & 64 & 3.8\% & -0.1 & 4.6\% & -0.2 & 1721824 & 66\% \\
    & 32 & 10.7\% & 0.3 & 12.2\% & -0.4 & 514848 & 90\% \\
    & 24 & 14.9\% & -0.4 & 18.9\% & -0.6 & 289944 & 94\% \\
    & 16 & 21.0\% & -0.5 & 25.6\% & -0.8 & 129168 & 97\% \\
    \hline
    \end{tabular}
    }
    \caption{BD-Rate and BD-PSNR for JPEG SCM using JPEG PCC as reference (for the full JPEG CTTC dataset).}
    \label{tab:5_rd_simp}
\end{table}

\begin{itemize}
    \item The three simplification approaches differ in the layers affected by the reduction in the number of channels, with obvious impact in the overall parameters’ reduction, which varies from 15\% (reducing only the number of channels of the latent representation) up to 97\% (only achievable by reducing the number of channels in all AE and VAE layers).
    \item The three simplification approaches have a significantly different impact on the RD performance: while reducing the number of latent channels and the number of latent and hyper latent channels has a lower impact in RD performance (below 2\% for all tested options), reducing the number of channels in all model layers can result in losses of up to 25\% in BD-Rate.
    \item Overall, the best model simplification approach is the reduction of the number of latent and hyper latent channels, which offers rather high complexity reduction ratios (between 42\% and 56\%), showing even slight BD-Rate gains of up to 1.7\% for PSNR D1.
    \item Considering the trade-offs offered by all studied approaches, the selected lightweight model (marked in bold in Table \ref{tab:5_rd_simp}) uses a number of latent and hyper latent channels equal to 16, since it achieves the best RD performance for a very significant 56\% reduction on the number of model parameters.
\end{itemize}

The small average RD performance gains for some of the simplified geometry coding models indicate that they are more efficient in extracting highly relevant features from a compression perspective than the original JPEG PCC coding model. The resulting latents provide a more compact representation without losing any of the relevant information required for high quality signal reconstruction. These results also show that JPEG PCC uses a percentage of the overall rate that is not cost-efficient in RD terms, notably many zero-latents which despite their small rate cost are effectively irrelevant in terms of quality contribution. It is important to remember that the parameter reduction percentage shown in Table \ref{tab:5_rd_simp} refers only to the PC geometry coding model, i.e., JPEG SCM codec. When the new SR-enabled coding model is considered, the overall complexity reduction offered by the SR-EPCC codec increases to around 70\% still with some small average compression gains.

\section{Conclusions}
\label{sec6}

Point cloud-based applications are becoming increasingly popular and impactful, critically needing efficient compression solutions due to the large number of points required to offer rich and immersive experiences. As the first learning-based coding standard for PCs, the JPEG PCC standard is opening a new era for 3D visual representation technology. This paper proposes a solution to significantly reduce the JPEG PCC complexity at no compression penalty, what is critical for the wide deployment of this type of coding standard, notably in complexity constrained consumption environments. The key novelties involve using compressed domain SR models, the first of its kind, and a simplified coding model. Future work will involve the study of the performance offered by this coding solution for computer vision tasks directly processing the compressed bitstreams without decoding.

\section{Acknowledgment}

This work was supported by Fundação para a Ciência e Tecnologia, I.P. (FCT, Funder ID = 50110000187), under the project with reference UIDB/50008/2020 (DOI: 10.54499/UIDB/50008/2020) and the project with reference PTDC/EEI-COM/1125/2021 entitled “Deep Learning-based Point Cloud Representation.”




\end{document}